\documentclass[runningheads]{llncs}
\usepackage[T1]{fontenc}
\usepackage{graphicx}
\usepackage{subcaption}
\usepackage{hyperref}
\usepackage{color}

\usepackage{amsmath,amsfonts}
\usepackage{bbm, dsfont}

\newcommand{\bdm}{\begin{displaymath}}
\newcommand{\edm}{\end{displaymath}}

\newcommand{\Rf}{{\mathbb R}}

\newcommand{\argmax}{\operatornamewithlimits{arg \, max}}

\newcommand{\udots}{
  \mathinner {\mkern 1mu\raise 1pt \vbox {\kern 7pt \hbox {.}}\mkern 2mu
  \raise 4pt \hbox {.}\mkern 2mu\raise 7pt \hbox {.}\mkern 1mu}}

\newcommand{\bi}{{\boldsymbol i}}
\newcommand{\bj}{{\boldsymbol j}}

\newcommand{\bn}{{\boldsymbol n}}

\newcommand{\bu}{{\boldsymbol u}}

\newcommand{\bx}{{\boldsymbol x}}

\newcommand{\Cl}[1]{\text{C}\ell_{#1, 0}}

\newcommand{\bff}{\mathbf{f}}
\newcommand{\R}{\mathcal{R}}

\begin{document}
\title{A Structurally Coherent Spatial Phase Estimate}
\titlerunning{A Structurally Coherent Spatial Phase Estimate}
%
\author{Brian Knight\inst{1}\orcidID{0000-0001-8049-4749} \and
Naoki Saito\inst{1}\orcidID{0000-0001-5234-4719}}
%
\authorrunning{B. Knight, N. Saito}
%
\institute{Department of Mathematics, University of California, Davis, CA 95616 USA}
%
\maketitle              
\begin{abstract}
The monogenic signal (MS) was introduced by Felsberg and Sommer 
\cite{felsberg_low-level_nodate}, and independently by Larkin \cite{larkin_vortex1_2001} 
under the name vortex operator. 
It is a two-dimensional (2D) analog of the well-known analytic signal, 
and allows for direct amplitude and phase demodulation of (amplitude and phase) modulated images 
so long as the signal is intrinsically one-dimensional (i1D). 
Felsberg’s PhD dissertation also introduced the structure multivector (SMV), 
a model allowing for intrinsically 2D (i2D) structure. 
While the monogenic signal has become a well-known tool in the image processing community, 
the SMV is little used, 
although even in the case of i1D signals it provides a more robust orientation estimation than the MS. 
We argue the SMV is more suitable in standard i1D image feature extraction due to the this 
improvement, and extend the steerable wavelet frames of Held et al. 
\cite{held_steerable_wavelets_2010} to accommodate the additional features of the SMV. 
We then propose a novel quality map based on local orientation variance which values structurally 
coherent patches. This yields a multiscale 
phase estimate which performs well even when signal to noise ratio (SNR) is $\le 1$. 
The performance is evaluated on several synthetic phase estimation tasks as well as on a fine-scale 
fingerprint registration task related to the 2D phase demodulation problem.

\end{abstract}

\section{Introduction}\label{sec:1_intro}
Many problems in imaging science 
rely on spatial phase measurements, e.g. 2D interferometry, 
interferometric SAR (InSAR), and require a preprocessing step 
to estimate the true spatial phase of an image or set of images
\cite{INSar_phase_estimation_rs15030613}.
Any improvement to this estimate will thus improve downstream analysis. 
A standard approach for estimating spatial phase of images is to use the phase of the 
\emph{monogenic signal} \cite{monogenic_phase_estimation_KULKARNI2019208} 
\cite{InSAR_https://doi.org/10.1155/2022/3086486}.
The first improvement to this estimate is to produce a multiscale monogenic phase estimate, 
see Kaseb et al. \cite{kaseb_phase_2019}, 
for instance, which makes use of isotropic wavelets \cite{held_steerable_wavelets_2010} \cite{unser_doi:10.1137/120866014}, 
and provides a robust phase estimate in the presence of image corruption.
\textbf{The main contributions of this article are:} 
\textbf{1)} to employ the \emph{structure multivector} (SMV) in place of the monogenic signal 
in order to extract a more robust feature set at any given scale; and
\textbf{2)} to define a novel quality measure at each scale, based on the features of the SMV, 
in order to determine the optimal local feature set around a given point in an image. 
We perform several experiments on synthetic images to showcase the application of the 
our multiscale phase estimation, and further solve a phase and amplitude demodulation problem in 2D
to display the utility of this estimate. Additionally, we use our multiscale phase estimate in 
order to solve a fine-scale fingerprint registration problem as described in \cite{cui_Deformable_8368301}.  
Lastly, we have provided a Julia module that includes the code needed to reproduce 
any figures and experiments shown in this paper, as well as standalone functions to perform our multiscale 
phase estimation: \hyperlink{https://gitlab.com/briancknight/SSVM2025}{https://gitlab.com/briancknight/SSVM2025}.

The rest of the article is organized as follows.
In Section \ref{sec:2_signal_models} we discuss the standard signal model used to describe fringe patterns, 
and define the monogenic signal (MS), steerable wavelets, and the structure multivector (SMV). 
Section \ref{sec:3_multiscale_features} describes the novel multiscale phase estimate 
derived from the SMV features at each scale, 
and Section \ref{sec:4_experiments} concludes with numerical experiments 
showcasing the improved phase estimation
as well as improved accuracy in 2D phase and amplitude demodulation tasks, 
including one used in fine-scale fingerprint registration.
Section \ref{sec:5_conclusion} provides a brief conclusion to the article.
In the additional appendix, Appendix \ref{sec:appendixA_SMV} provides more details 
in the construction of the structure multivector, and
Appendix \ref{sec:appendixB_orientation} provides an error analysis for the feature set of the SMV 
when the i2D signal model is violated. 

\section{Signal Models}\label{sec:2_signal_models}
\subsection{The i1D Signal Model and the Monogenic Signal}

In 1D signal processing, a real-valued signal can be extended to a complex valued signal via the Hilbert
transform, and this complex-valued extension can provide useful insight into the signal's local amplitude 
and frequency. In 2D, the monogenic signal is a quaternion valued signal that extends the real-valued 
2D signal via the Riesz transform,the appropriate 2D extension of the Hilbert transform. 
If our original signal obeys the signal model
\begin{equation}\label{eq:i1DSignal}
    f(\bx) = A(\bx) \cos(\bn(\bx)\cdot\bx),
    \quad \bx \in \Rf^2
\end{equation}
where $A(\bx) \ge 0$ is the local amplitude function, and $\bn(\bx)$ is the local orientation, 
and $\varphi(\bx) = \bn(\bx)\cdot\bx$ is the local phase function, 
and we assume $A(\bx)$ varies slowly with respect to the $\cos(\bn(\bx)\cdot\bx)$. 
We may also refer to $A$ as the \textit{local energy}, 
and the tuple $(\bn(\bx), \varphi(\bx))$ as the \textit{local structure}.

The \textit{monogenic signal} is defined as the quaternionic signal
\begin{equation*}
    \begin{split}
    f_M(\bx) &= f(\bx) + \bi \R_1f(\bx) +\bj \R_2f(\bx), 
    \end{split}
\end{equation*}
where 
\begin{equation*}
    \R_{k}f(\bx) = \int_{\Rf^2} \frac{x'_k}{2\pi\|\bx'\|^{\frac{3}{2}}} f(\bx - \bx')d x'_1 d x'_2,
\end{equation*}
is the Riesz transform for the $x_k$ direction in $\Rf^2$, and $\R f(\bx) = \bi \R_1f(\bx) +\bj \R_2f(\bx)$ is 
the (total) Riezs transform.
Larkin \cite{larkin_vortex1_2001} showed that for signals of the i1D model described above, 
the Riesz transform obeys the asymptotic Bedrosian Principle:
\begin{equation*}
    \begin{split}
    \R(A(\bx) \cos(\varphi(\bx)) &\approx A(\bx) \R(\cos(\varphi(\bx))),
    \end{split}
\end{equation*}
and, further, that 
\begin{equation*}
    \begin{split}
    \R(\cos(\varphi(\bx))) \approx \bn(\bx) \sin(\varphi(\bx)),
    \end{split}
\end{equation*}
In the case of $\varphi(\bx) = 2\pi\omega \bn \cdot \bx$ for some unit vector $\bn$ and frequency $\omega$,
i.e., a plane wave, so long as $\hat{A}(\bu) = 0$ for any $\bu$ such that $|\bu| > \omega$, 
the Riesz transform performs the correct quadrature shift for phase estimation of i1D signals in two dimensions.

To be clear, given a signal $f$ of the form described in \eqref{eq:i1DSignal}, 
the monogenic signal of $f$ is given by
\begin{equation*}
    f_M(\bx) = A(\bx) \left(\cos(\varphi(\bx)) + \bn(\bx) \sin(\varphi(\bx))\right),
\end{equation*}
thus we can recover the local amplitude, local orientation, and local phase via 
\begin{equation*}
    A(\bx) = |f_M(\bx)|,\quad
    \bn(\bx) = \frac{\R f(\bx)}{|\R f(\bx)|},\quad
    \varphi(\bx) = \arctan\left(|\R f(\bx)|, f(\bx)\right).
\end{equation*}

This feature set is considered to be a \textit{split of identity}, in that it separates a signal into 
independent local features. Specifically, the local structure is invariant to scaling of the local energy,
and the local energy is invariant to phase shifts in the local structure. 
This allows for useful image processing steps, such as equalization of 
brightness \cite{held_steerable_wavelets_2010}, or, as we discuss later, 
phase modulation and demodulation.

Similarly, we can use local amplitude information in order to determine important features. 
This approach is particularly useful when the monogenic signal is paired with an isotropic 
wavelet decomposition \cite{kaseb_phase_2019}, which we outline in the next section.

\subsection{Steerable Wavelet Frames using Riesz Transforms}
The Riesz transform $\R$ is isotropic, meaning that if $R_{\theta}$ is a 2D rotation matrix,
then $\R R_{\theta}f(\bx) = R_{\theta} \R f(\bx)$. It also commutes with shifts and dilations 
so that applying any one of these operations to the signal $f$ can be done before
or after computing the monogenic extension. As far as wavelet analysis is concerned, this allows
us to construct monogenic wavelets simply by constructing a real isotropic wavelet 
and then computing its Riesz transform.

Based on this property, Held et al. \cite{held_steerable_wavelets_2010} construct steerable wavelet 
frames for $n$D signals which, for a given image $f \in \Rf^{2^M \times 2^M}$, 
yields the decomposition into $M\cdot K$ scales, 
where $M$ is the number of dyadic scales, and $K$ the number of subscales used for each dyadic scale. 
We denote the components of the decomposition by
$d_{j,s}(\bx)$ for $j = 1, \dots K$, $s = M', \dots, M$, where $M'\ge1$ is the smallest 
dyadic scale used (typically $M' = 3$), and an approximation 
component $a_M$ which contains any remaining low-frequency information. 
Naturally, we can extend each scale via the Riesz transform to obtain:
\begin{equation*}
        (d_{j,s}(\bx))_{M} = A_{j,s}(\bx)\left[\cos(\varphi_{j,s}(\bx)) + \bn_{j,s}(\bx) \sin(\varphi_{j,s}(\bx))\right]
\end{equation*}

\begin{figure}[ht] 
    \centering
    \begin{subfigure}{0.45\textwidth}
    \centering
        \begin{subfigure}{0.3\textwidth}
            \renewcommand\thesubfigure{\alph{subfigure}1}
            \includegraphics[width=\textwidth]{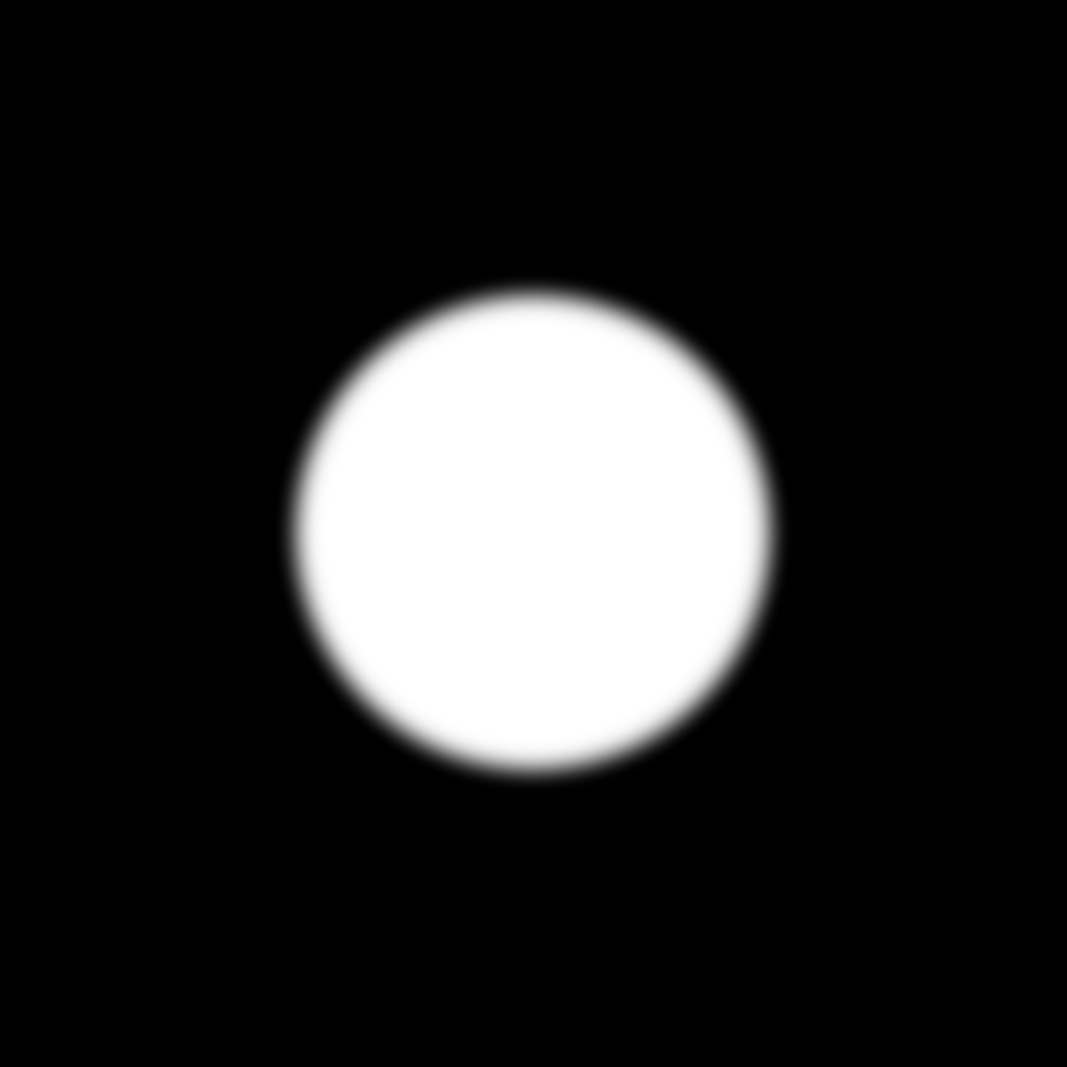}
            \centering
            \caption{}
        \end{subfigure}
        \begin{subfigure}{0.3\textwidth}
            \addtocounter{subfigure}{-1}
            \renewcommand\thesubfigure{\alph{subfigure}2}
            \centering
            \includegraphics[width=\textwidth]{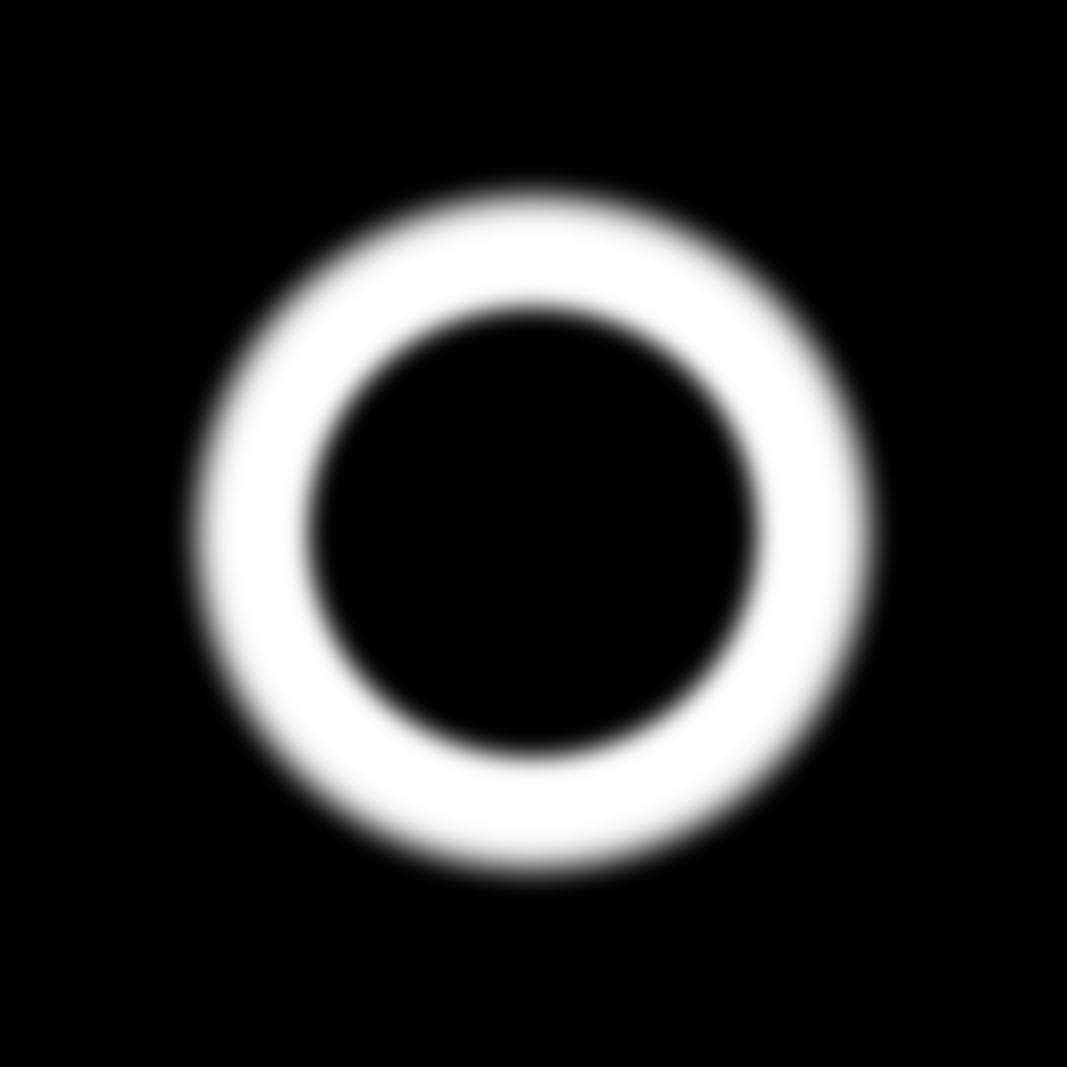}
            \caption{}
        \end{subfigure}
        \begin{subfigure}{0.3\textwidth}
            \addtocounter{subfigure}{-1}
            \renewcommand\thesubfigure{\alph{subfigure}3}
            \centering
            \includegraphics[width=\textwidth]{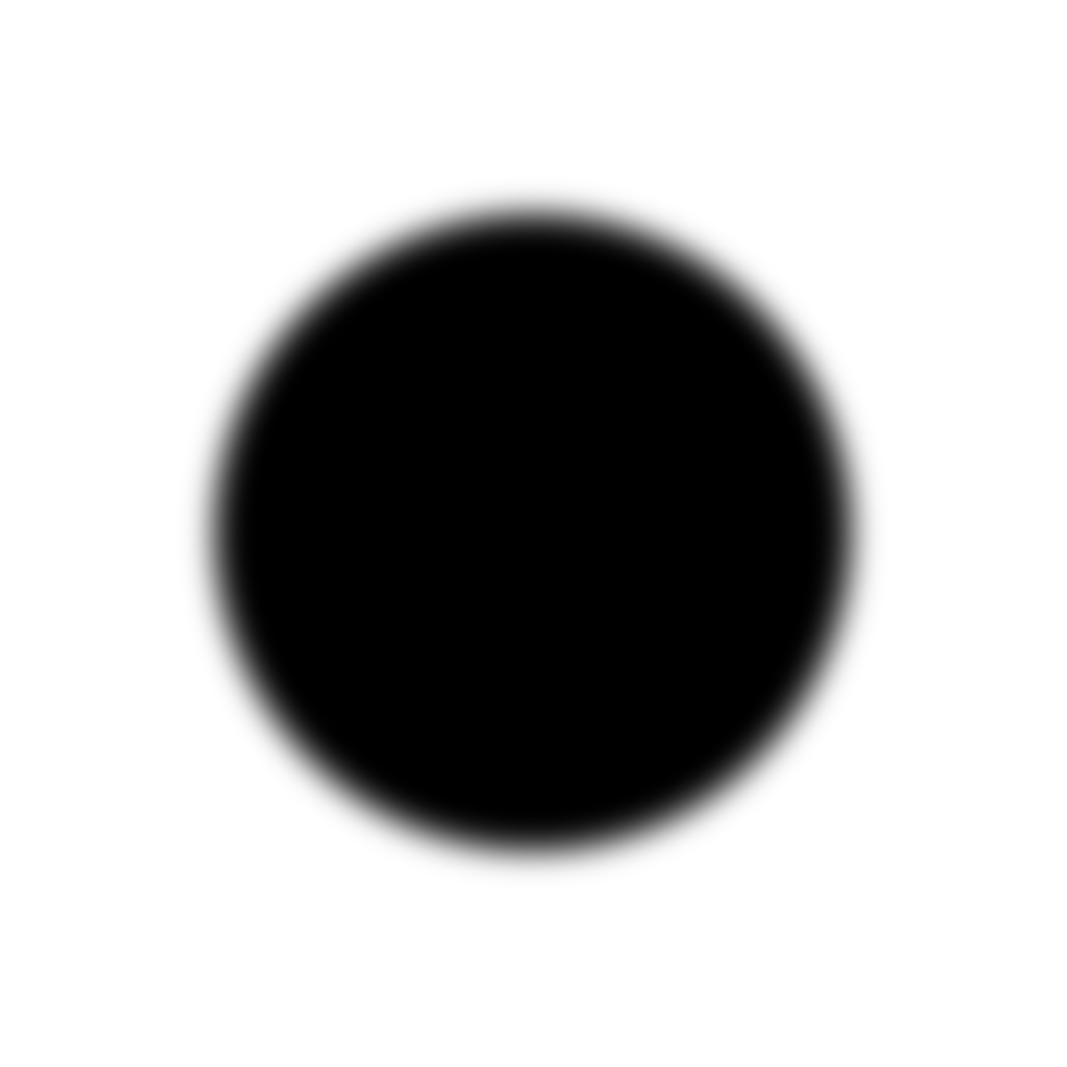}
            \caption{}
        \end{subfigure}
        \addtocounter{subfigure}{-1}
        \begin{subfigure}{0.3\textwidth}
            \renewcommand\thesubfigure{\alph{subfigure}4}
            \includegraphics[width=\textwidth]{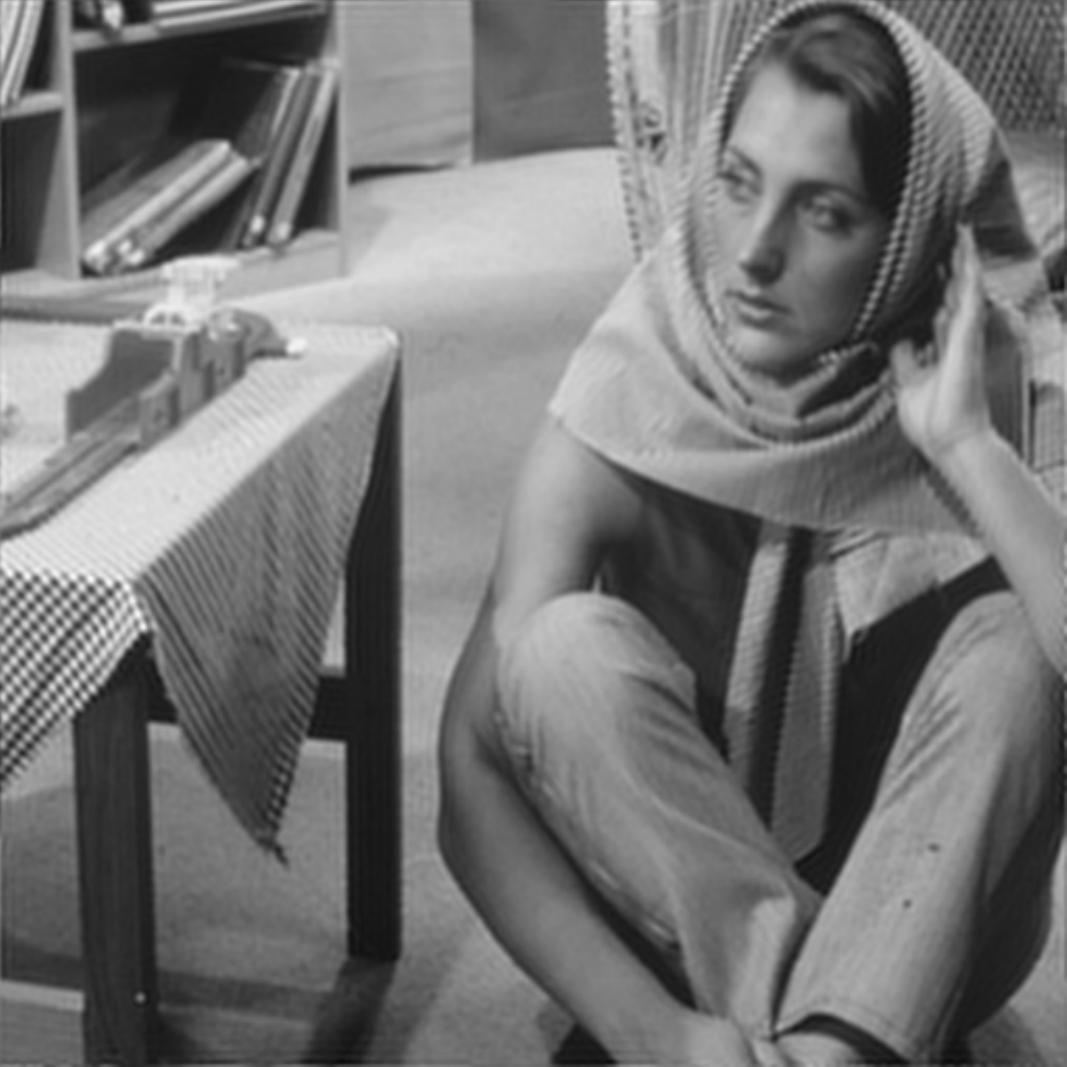}
            \centering
            \caption{}
        \end{subfigure}
        \begin{subfigure}{0.3\textwidth}
            \addtocounter{subfigure}{-1}
            \renewcommand\thesubfigure{\alph{subfigure}5}
            \centering
            \includegraphics[width=\textwidth]{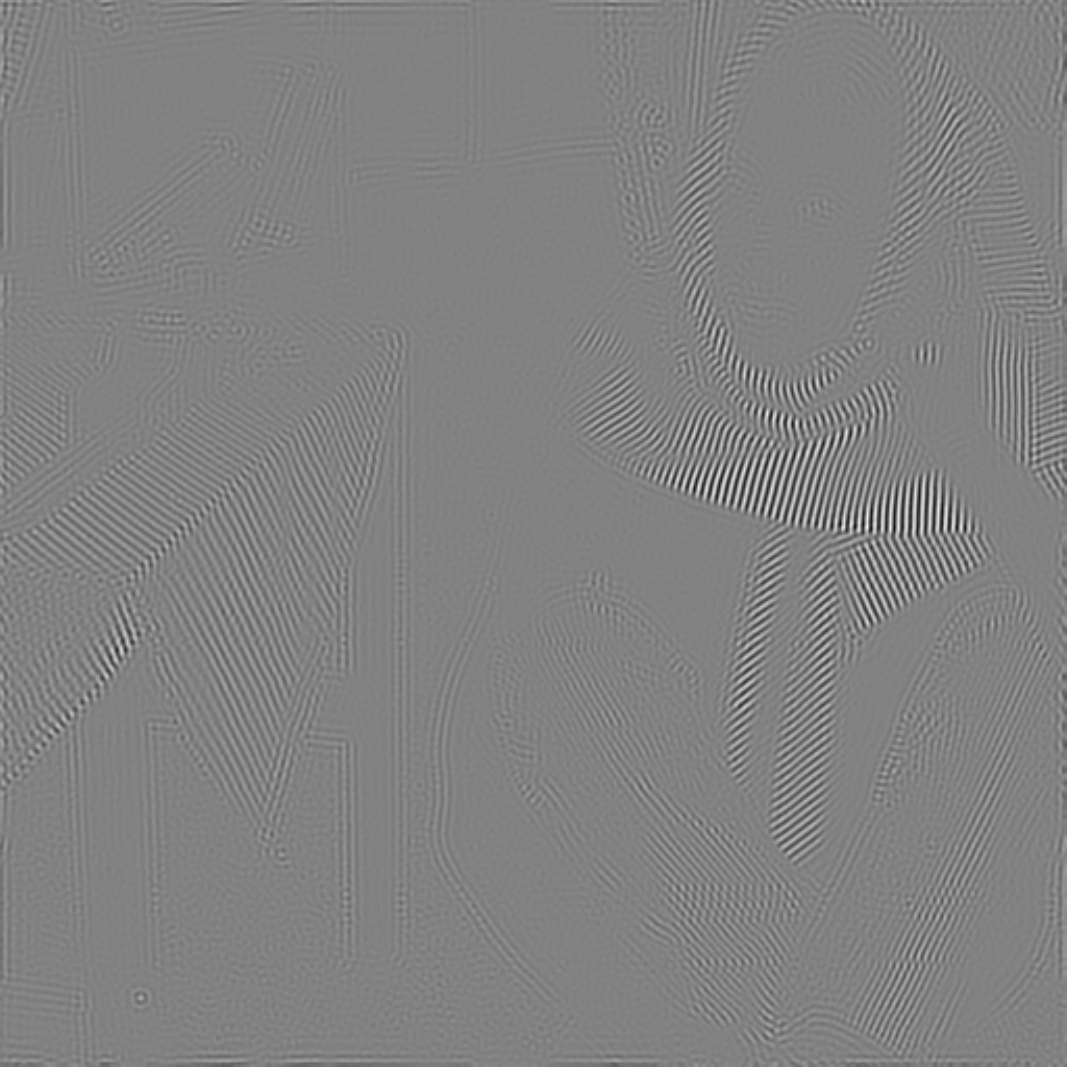}
            \caption{}
        \end{subfigure}
        \begin{subfigure}{0.3\textwidth}
            \addtocounter{subfigure}{-1}
            \renewcommand\thesubfigure{\alph{subfigure}6}
            \centering
            \includegraphics[width=\textwidth]{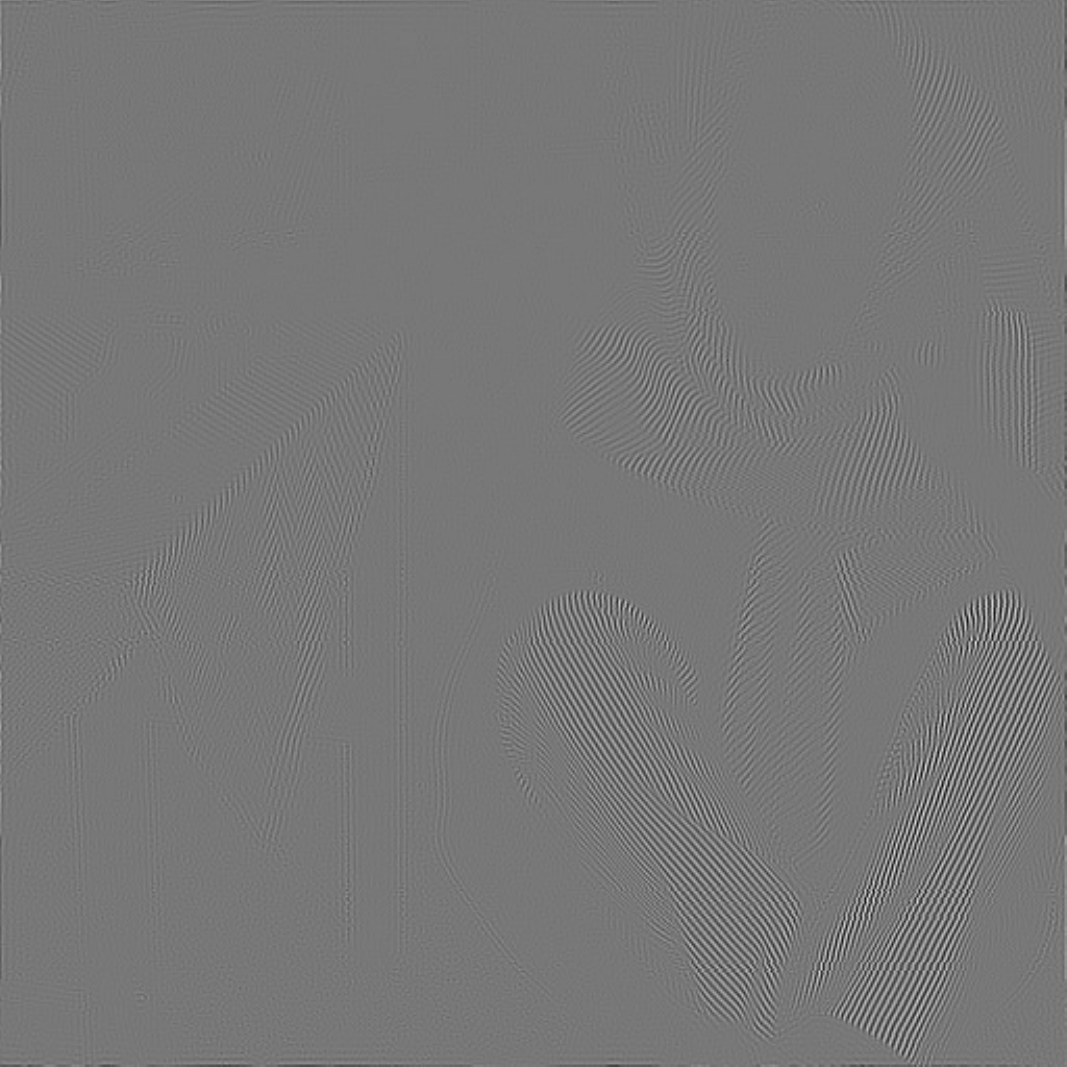}
            \caption{}
        \end{subfigure}
    \end{subfigure}
    ~
    \begin{subfigure}{0.45\textwidth}
        \centering
        \begin{subfigure}{0.3\textwidth}
            \renewcommand\thesubfigure{\alph{subfigure}1}
            \includegraphics[width=\textwidth]{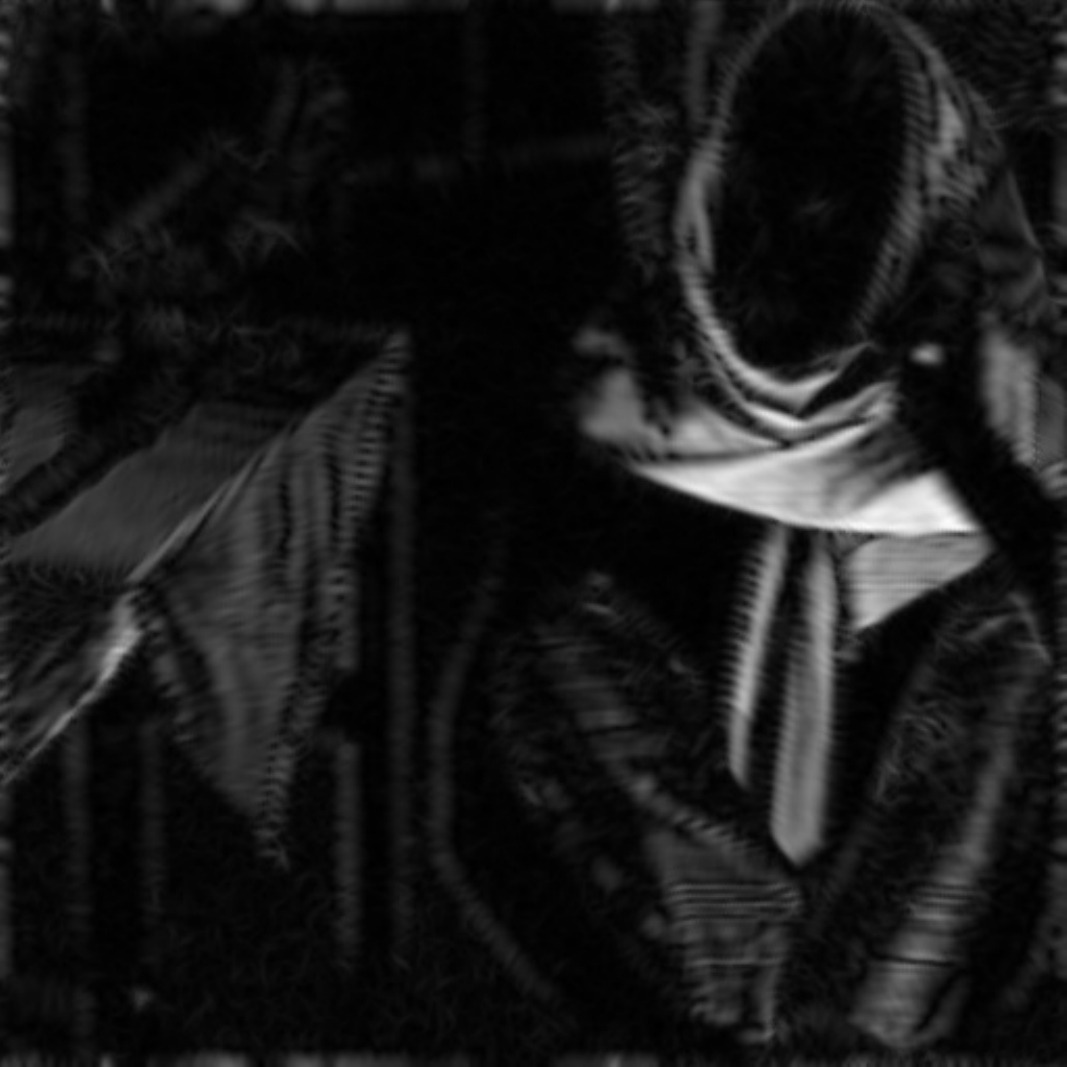}
            \centering
            \caption{}
        \end{subfigure}
        \begin{subfigure}{0.3\textwidth}
            \addtocounter{subfigure}{-1}
            \renewcommand\thesubfigure{\alph{subfigure}2}
            \centering
            \includegraphics[width=\textwidth]{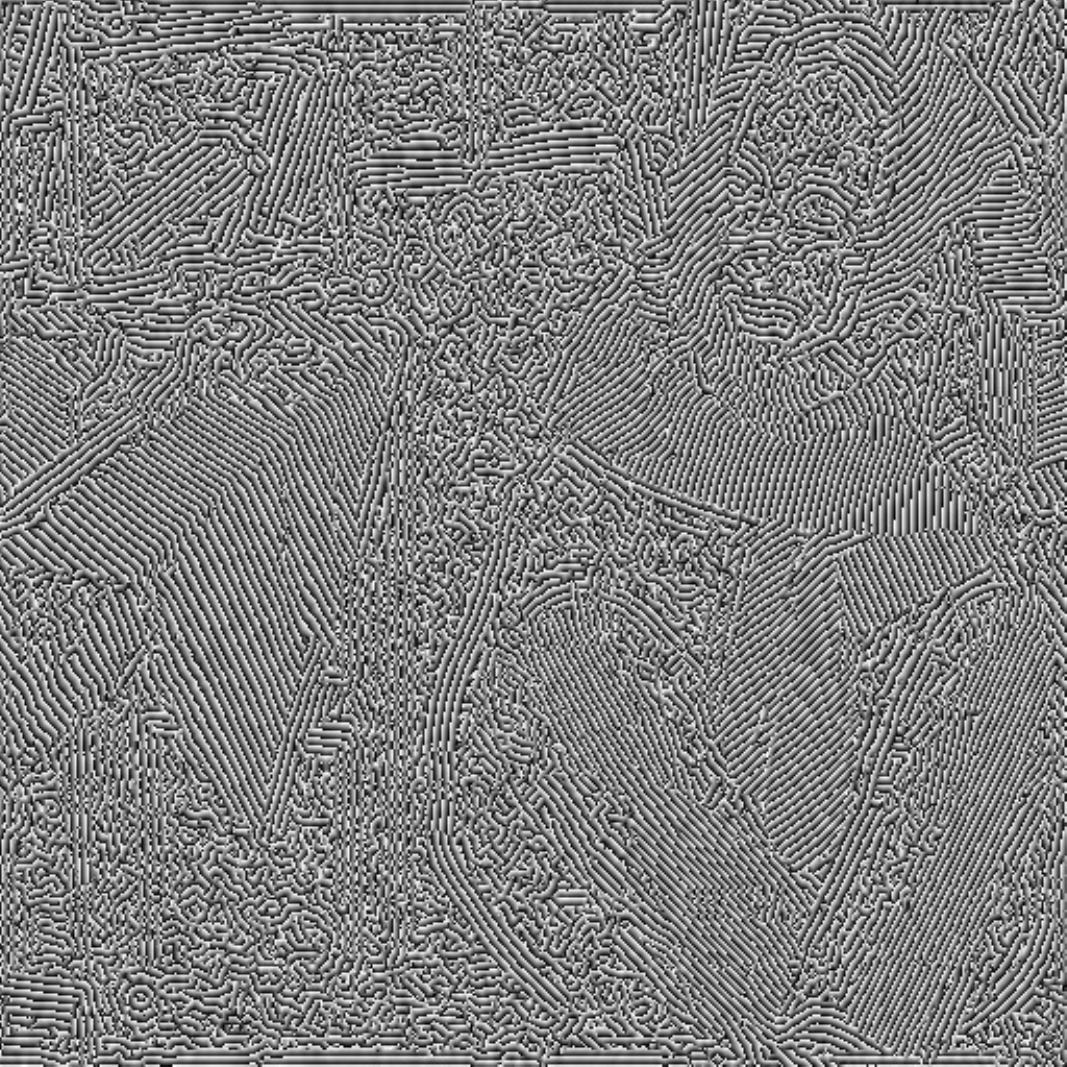}
            \caption{}
        \end{subfigure}
        \begin{subfigure}{0.3\textwidth}
            \addtocounter{subfigure}{-1}
            \renewcommand\thesubfigure{\alph{subfigure}3}
            \centering
            \includegraphics[width=\textwidth]{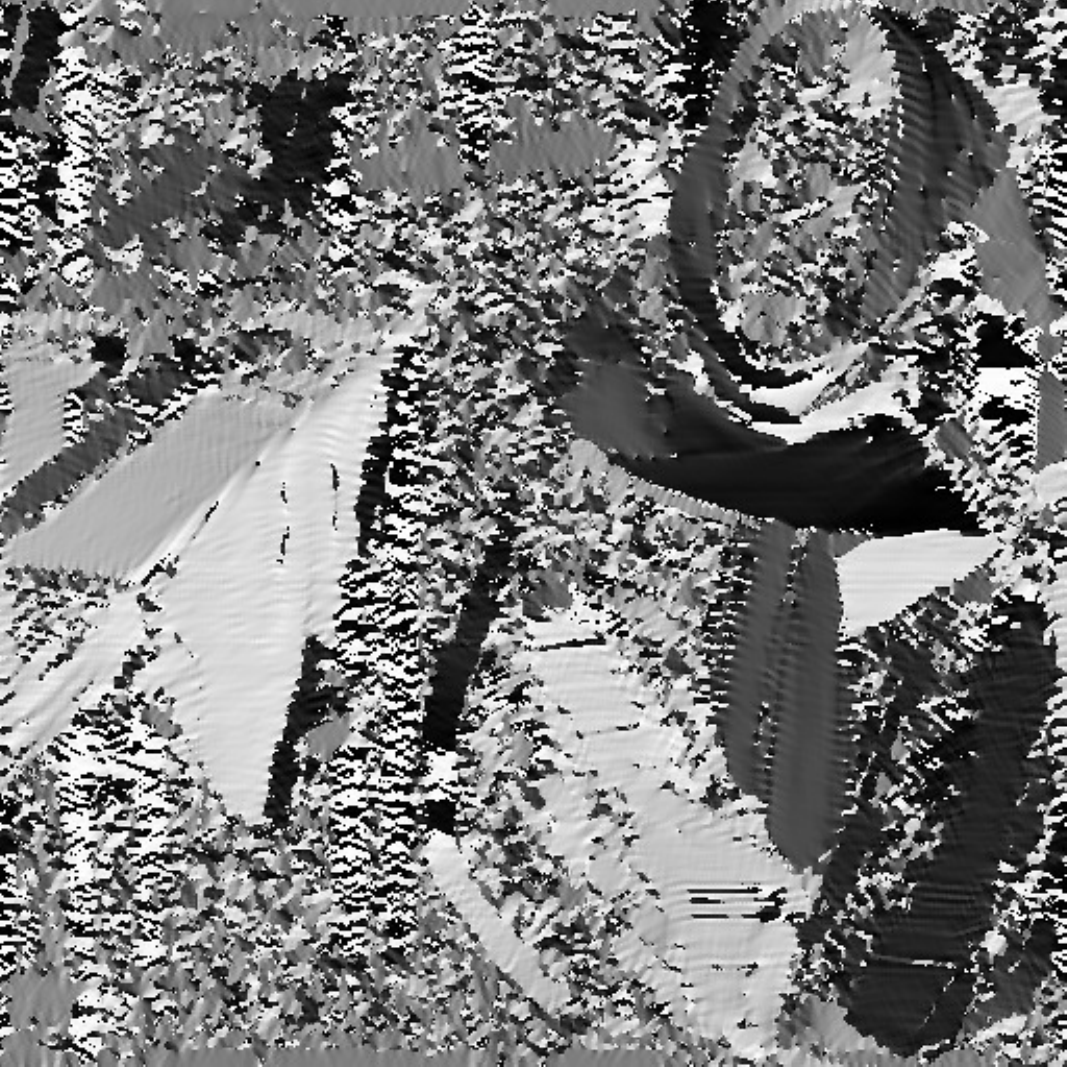}
            \caption{}
        \end{subfigure}
        \addtocounter{subfigure}{-1}
        \begin{subfigure}{0.3\textwidth}
            \renewcommand\thesubfigure{\alph{subfigure}4}
            \includegraphics[width=\textwidth]{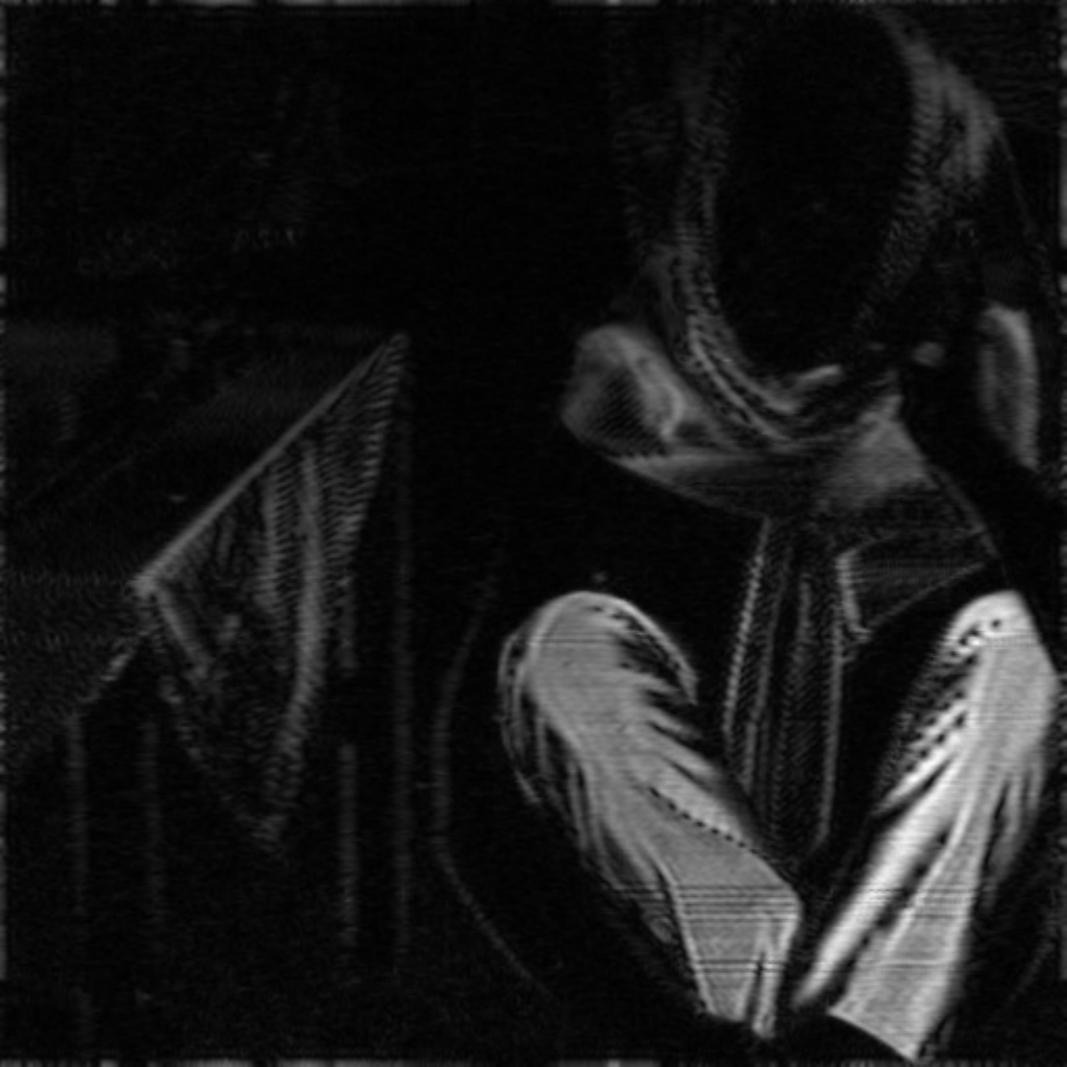}
            \centering
            \caption{}
        \end{subfigure}
        \begin{subfigure}{0.3\textwidth}
            \addtocounter{subfigure}{-1}
            \renewcommand\thesubfigure{\alph{subfigure}5}
            \centering
            \includegraphics[width=\textwidth]{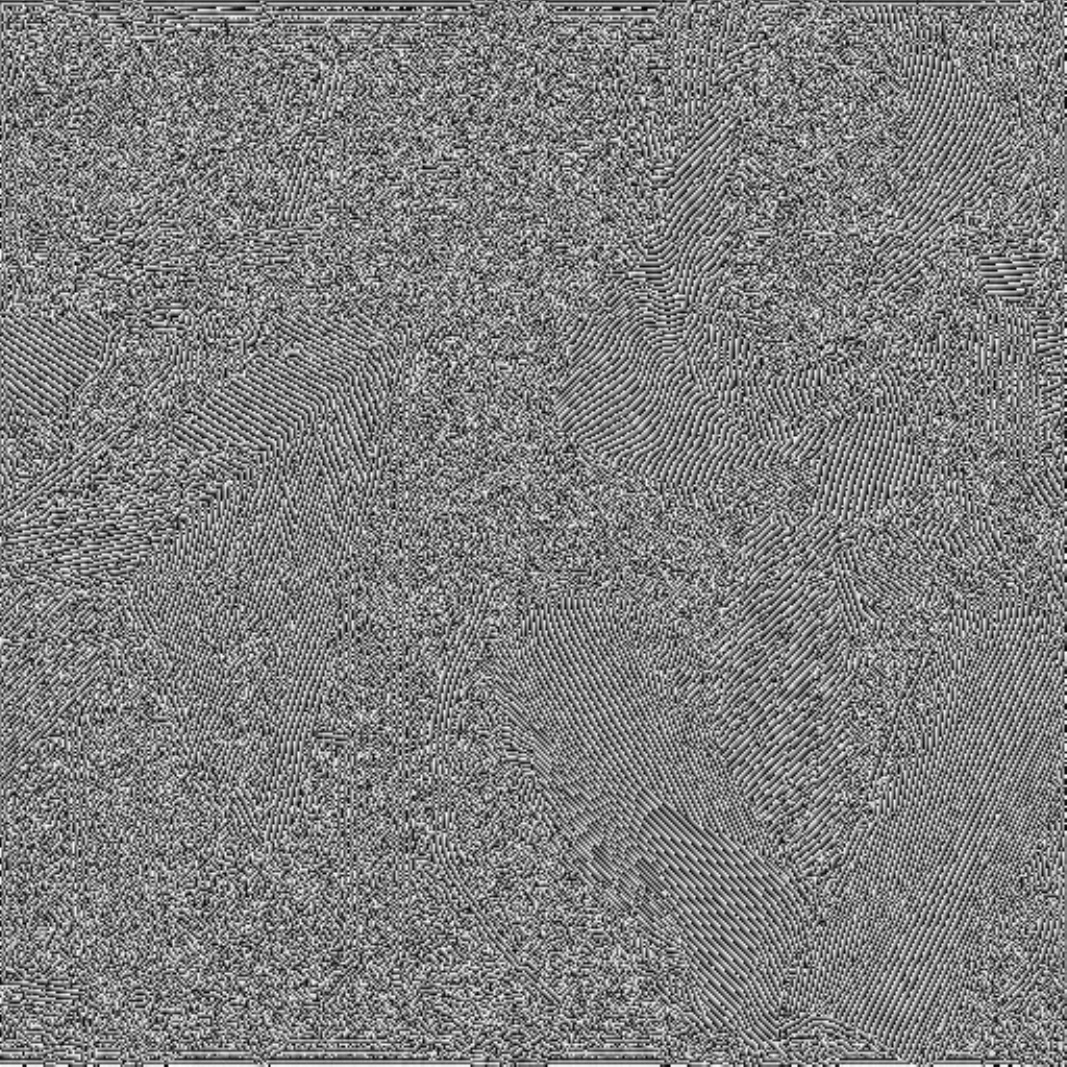}
            \caption{}
        \end{subfigure}
        \begin{subfigure}{0.3\textwidth}
            \addtocounter{subfigure}{-1}
            \renewcommand\thesubfigure{\alph{subfigure}6}
            \centering
            \includegraphics[width=\textwidth]{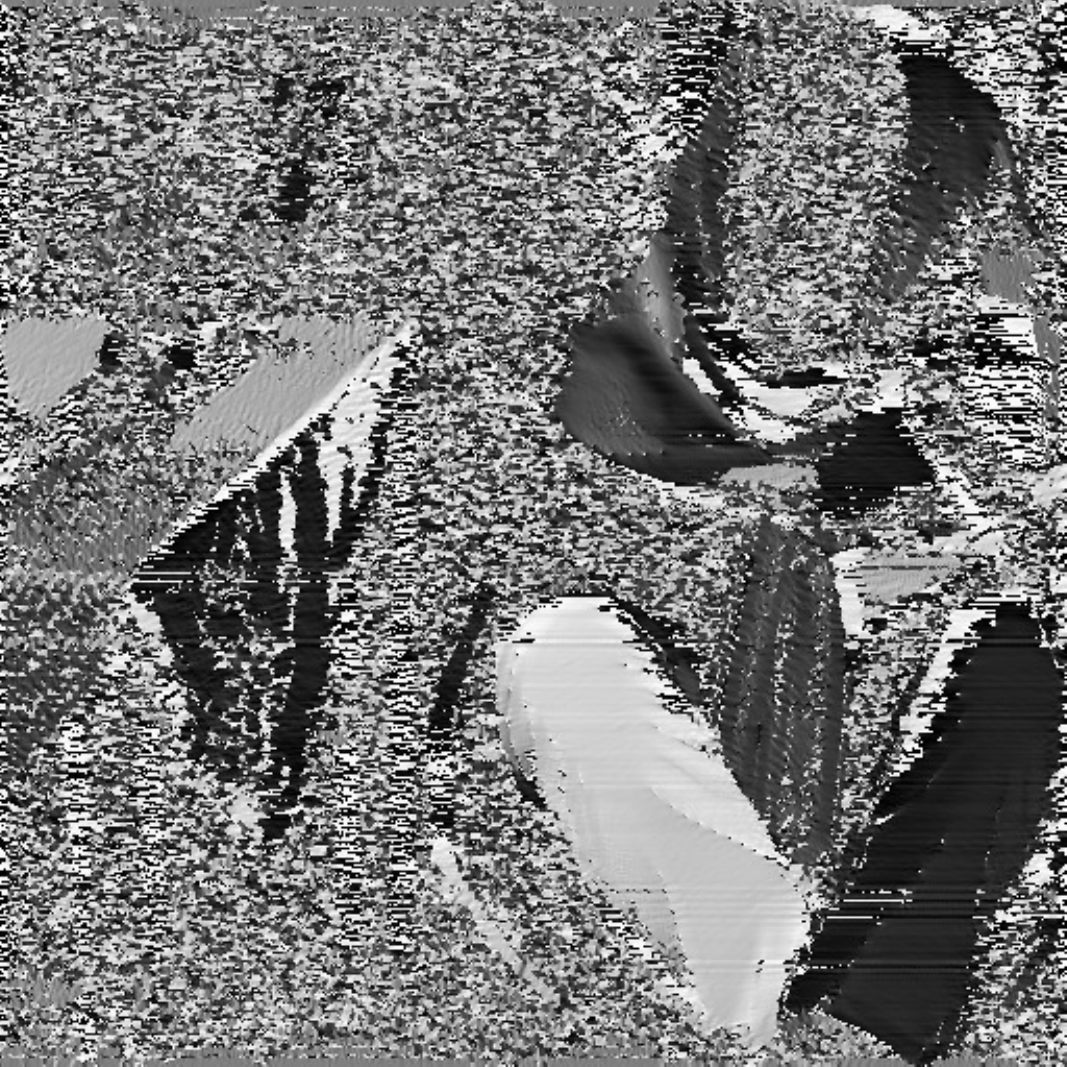}
            \caption{}
        \end{subfigure}
    \end{subfigure}%
    \caption{Left: Isotropic wavelet frames, (a1) low-pass $h_{1,1}$, (a2) band-pass $h_{1,2}$, 
    (a3) high-pass $h_{1,3}$. Applied to Barbara: (a4) low-pass, (a5) band-pass, (a6) high-pass. 
    Right: IAP representation of isotropic wavelet scales, (b1) $A_{1,2}(\bx)$, (b2) $\Phi_{1,2}(\bx)$, 
    (b3) $\theta_{1,2}(\bx)$,
    (b4) $A_{1,3}(\bx)$, (b5) $\Phi_{1,3}(\bx)$, (b6) $\theta_{1,3}(\bx)$}
    \label{fig:steerable_wvlts_barbara}
\end{figure}

If $K = 1$, each dyadic scale is decomposed via complementary high and low pass filters.
For $K > 1$, the dyadic scale is decomposed into $K-1$ band-pass components, 
and a high and low frequency component. 
The specific construction can be found used in \cite{held_steerable_wavelets_2010}. 
Figure~\ref{fig:steerable_wvlts_barbara}a depicts the filters for $K = 2$,
while Figure~\ref{fig:steerable_wvlts_barbara}b shows the instantaneous amplitude and phase (IAP) representation 
of the high frequency and band-pass component provided by the structure multivector
which we discuss in the next section.

\subsection{The i2D Signal Model and the Structure Multivector}

Two-dimensional signals can, of course, vary in two or more orientations in any given 
local patch, and further work has been done using hypercomplex signal processing that can deal with these cases. 
One extension, the \textit{structure multivector} (SMV), 
was introduced along with the monogenic signal in the PhD dissertation of Felsberg
\cite{felsberg_low-level_nodate}. It is designed to deal with signals of the form
\begin{equation}\label{eq:i2DSignal}
    f(\bx) = f_1(\bn(\bx)\cdot\bx) + f_2(\bn(\bx)^{\perp} \cdot \bx),
\end{equation}
These are intrinsically 2D (i2D) signals with two orientations in each local patch that are 
orthogonal to one another.

The features of the SMV will essentially be that of two monogenic signals, and to accommodate these 
additional features the SMV lives in a larger dimensional Clifford algebra, $\Cl{3}$ which 
subsumes the quaternions. This algebra is generated by the orthonormal basis $e_1, e_2, e_3$ satisfying the 
relations $e_i e_j + e_j e_i = 2\delta_{ij}$, and consists of $2^3 = 8$ elements:
$1, e_1, e_2, e_3, e_{12}, e_{23}, e_{31}, e_{123}$.
In general we denote the product 
$e_{i_{1}}e_{i_{2}}\cdots e_{i_{n}} := e_{i_{1}i_{2} \cdots i_{n}}$. 
For more information on Clifford algebras and the construction of the SMV see \cite{delanghe_clifford_2001}, 
\cite{felsberg_low-level_nodate} and Appendix \ref{sec:appendixA_SMV}.
We only give here the minimum details needed to construct the SMV.
We consider an image of the form $\bff:\Rf^2 \to e_3\Rf$,
$\bff(\bx) = f(x,y)e_3.$

The corresponding structure multivector (SMV) is given by 

\begin{equation}\label{eq:SMV}
    \begin{split}
        M_S(\bx) &=  \left[\bff(\bx) + (h_2^1 * \bff)(\bx)\right] + 
        e_3 \left[(h_2^2*\bff)(\bx) + (h_2^3 * \bff)(\bx)\right]\\
        &= M_0 + M_1 e_1 + M_2 e_2 + M_3 e_3 + 
        M_{23}e_{23} + M_{31} e_{31} + M_{12} e_{12}.
    \end{split}
\end{equation}

The explicit definitions of these functions are given below, 
where $\bx = x e_1 + y e_2$:
\begin{align*}
        M_{1} &= \frac{x}{2\pi|\bx|^3} * f(\bx), &
        M_{2} &= \frac{y}{2\pi|\bx|^3} * f(\bx), &
        M_{3} &= f(\bx),\\
        M_{23} &= \frac{3(3x^2y - y^3)}{2\pi|\bx|^5}*f(\bx), &
        M_{31} &= \frac{3(3xy^2 - x^3)}{2\pi|\bx|^5}*f(\bx), &&\\
        M_{0} &= \frac{-2(x^2 - y^2)}{2\pi|\bx|^4} * f(\bx), &
        M_{12} &= \frac{-4xy}{2\pi|\bx|^4} * f(\bx). &&
\end{align*}
Here $h_{2}^1 = \mathcal{R}$ denotes the Riesz transform, $h_2^3$ is the composition of $h_2^2$ and $h_{2}^1,$, 
where $H_2^2(\bu) = \frac{(u^2 - v^2) + 2uv e_{12}}{\bu^2}$
is the Fourier transform of $h^2_2$. $H^2_2$ responds only to even signals, 
and any two perpendicular vectors $\bn$ and $\bn^{\perp}$ become antiparallel after action by $H_2^2$, 
which means that an even signal according to the \eqref{eq:i2DSignal} will yield a response to $H^2_2$ 
whose argument is precisely twice that of the main orientation of $\bn$. 

Specifically, we can calculate the orientation $\bn$ given a signal of 
the form $f(\bx) = A\cos(\bn\cdot\bx) + B\cos(\bn^{\perp}\cdot\bx)$ directly from this response. 
To handle odd structures, we finally take the Riesz 
transform of $h_2^2$ to yield $h_3^2$. 
The product of the Riesz response and the response of the third order harmonic estimates this same orientation,
but is better suited for odd structures, 
hence the average of these two arguments provides a robust orientation estimate of the structure multivector, 
as given in Felsberg's dissertation \cite{felsberg_low-level_nodate}:

\begin{equation}\label{eq:SMVOrientationEst}
    \theta_e = \frac14 \arg \left[(M_0 + M_{12} I_2)^2 + (M_1 + M_2 I_2)(M_{31} - M_{23}I_2)\right],
\end{equation}
where $I_2 = e_{12}$ acts as the imaginary unit $\bi$.

In \cite{felsberg_low-level_nodate} the author shows that the extended signal model provides a more robust orientation estimate 
than that of the monogenic signal. This is further confirmed in \cite{sinusoidal_image_model}. 
In addition to these facts, we show that: 1) the feature set of the SMV is robust even to i2D signals which violate 
the orthogonality constraint; and 2) if one of the local i1D signal dominates the local energy, 
then we can estimate the corresponding orientation well even in the case of large deviation from this 
constraint. See Appendix~\ref{sec:appendixB_orientation} for details. 

With this orientation estimate it is then possible to construct 
a pair of local angular filters that decompose a signal $f$ into two i1D signals, 
which then yields two local amplitudes, two local orientations, 
and two local phases that can be used for further processing. Again, see \cite{felsberg_low-level_nodate} or 
Appendix \ref{sec:appendixA_SMV} for full details. 
The output of the local angular filtering is two complex i1D signals which we denote by $F_1(\bx)$ and $F_2(\bx)$.

The full feature set of the SMV then is given by the local orientation estimate in \eqref{eq:SMVOrientationEst} and:
\begin{equation}
        A_i(\bx) =  |F_i(\bx)|,\quad \phi_i(\bx) = \arg|F_i(\bx)|,
\end{equation}
for $i = 1, 2$. 
At each location $\bx$, we choose the main signal by selecting the pair with the largest local amplitude. 
This selection is given by the dominance index $d(\bx) = \argmax_{1,2} \{A_1(\bx), A_2(\bx)\}$, 
so that we have a major and minor IAP representation given by:

\begin{equation*}
    A(\bx) =  A_{d(\bx)}(\bx), \quad \Phi(\bx) = \phi_{d(\bx)}(\bx),\quad
a(\bx) =  A_{3 - d(\bx)}(\bx), \quad \phi(\bx) = \phi_{3 - d(\bx)}(\bx).
\end{equation*}
Here the capital $A$ and $\Phi$ denote the dominant, or major, local i1D signal. 
Figure~\ref{fig:steerable_wvlts_barbara}b depicts the major IAP representation of two scales of Barbara.

\section{Multiscale Feature Estimation}\label{sec:3_multiscale_features}

\subsection{A Multiscale Phase Estimate Using the Structure Multivector}
Here we extend the multiscale phase extraction algorithm to use the feature set of the structure multivector. 
Let $f$ be the given signal
and $(f^{(k)})_{k=1}^{K}$ be the isotropic wavelet decomposition of $f$, and $A^{(k)}, 
\Phi^{(k)}, a^{(k)},$ and $\phi^{(k)}$ for $k = 1, \dots, K$ denote the  major amplitude, 
major phase, minor amplitude, and minor phase of $f^{(k)}$ respectively. 

If we let $\mathcal{Q}(\bx) \ge 0$ be a \textit{local quality function} which may depend on 
any aspect of the local multiscale feature set (here provided by steerable wavelets and the SMV), we can
define the \textit{local scale} to be given by
\begin{equation*}
    k^{\mathcal{Q}}(\bx) = \arg \max_{k} \{\mathcal{Q}^{(k)}(\bx)\},
\end{equation*}
where $\mathcal{Q}^{(k)}$ is the quality function applied to the multiscale features at scale $f^{(k)}$. 
The corresponding local \textit{multiscale features}
\begin{align*}
        A_{\mathcal{Q}}(\bx) &= A^{k^{\mathcal{Q}}(\bx)}(\bx), & \Phi_{\mathcal{Q}}(\bx) &= \Phi^{k^{\mathcal{Q}}(\bx)}(\bx), 
        & \theta_{\mathcal{Q}}(\bx) &=  \theta^{k^{\mathcal{Q}}(\bx)}(\bx),\\
        a_{\mathcal{Q}}(\bx) &= a^{k^{\mathcal{Q}}(\bx)}(\bx), & \phi_{\mathcal{Q}}(\bx) &= \phi^{k^{\mathcal{Q}}(\bx)}(\bx). &&
\end{align*}
In \cite{kaseb_phase_2019} $\mathcal{Q}^{(k)}(\bx)$ is defined to be the 
amplitude of the monogenic signal at that scale. The analog in this paper 
is the major amplitude of the SMV at each scale, which we call the \textit{local amplitude quality}.
The idea is to choose the scale with the maximum local energy; 
if the underlying signal in question is some sort of well structured fringe pattern, 
such as an interferogram or fingerprint,
this scale should correspond to the true spatial phase to be estimated
and the multiscale feature set should be robust to noise or local signal corruption 
when the signal to noise ratio (SNR) is sufficiently well behaved.

When the local SNR is close to 1, however, the scale with the dominant amplitude is likely to be the noise itself. 
Still, it may be assumed that the signal corruption does not contain 
coherent structural information, and so we submit that a local quality metric which 
makes use of the local structural information should enable good phase estimation even 
when SNR $\le 1$. 

\begin{figure}[ht] 
    \setlength{\belowcaptionskip}{-1.3\baselineskip}
    \centering
    \begin{subfigure}{0.19\textwidth}
        \includegraphics[width=\textwidth]{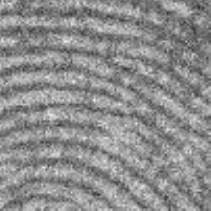}
        \caption*{}
    \end{subfigure}
    \begin{subfigure}{0.19\textwidth}
        \includegraphics[width=\textwidth]{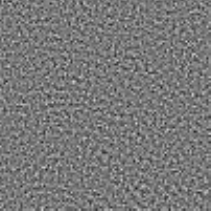}
        \caption*{}
    \end{subfigure}
    \begin{subfigure}{0.19\textwidth}
        \includegraphics[width=\textwidth]{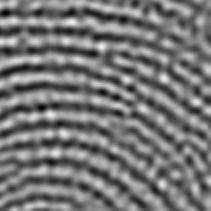}
        \caption*{}
    \end{subfigure}
    \begin{subfigure}{0.19\textwidth}
        \includegraphics[width=\textwidth]{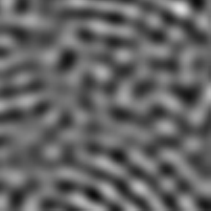}
        \caption*{}
    \end{subfigure}
        \begin{subfigure}{0.19\textwidth}
        \includegraphics[width=\textwidth]{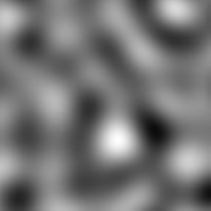}
        \caption*{}
    \end{subfigure}
    \par
    \begin{subfigure}{0.19\textwidth}
        \includegraphics[width=\textwidth]{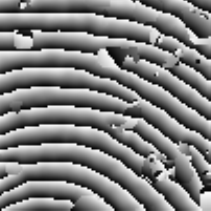}
        \caption*{}
    \end{subfigure}
    \begin{subfigure}{0.19\textwidth}
        \includegraphics[width=\textwidth]{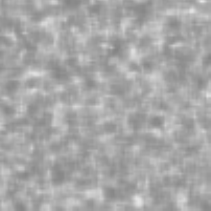}
        \caption*{}
    \end{subfigure}
    \begin{subfigure}{0.19\textwidth}
        \includegraphics[width=\textwidth]{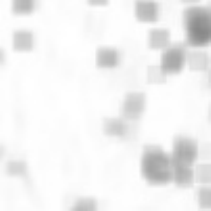}
        \caption*{}
    \end{subfigure}
    \begin{subfigure}{0.19\textwidth}
        \includegraphics[width=\textwidth]{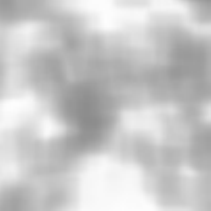}
        \caption*{}
    \end{subfigure}
    \begin{subfigure}{0.19\textwidth}
        \includegraphics[width=\textwidth]{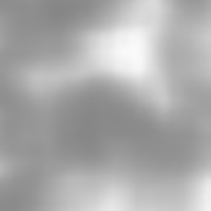}
        \caption*{}
    \end{subfigure}
    \par
    \begin{subfigure}{0.19\textwidth}
        \includegraphics[width=\textwidth]{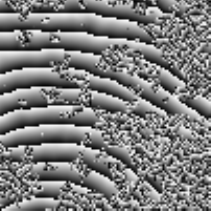}
        \caption*{}
    \end{subfigure}
    \begin{subfigure}{0.19\textwidth}
        \includegraphics[width=\textwidth]{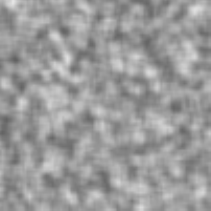}
        \caption*{}
    \end{subfigure}
    \begin{subfigure}{0.19\textwidth}
        \includegraphics[width=\textwidth]{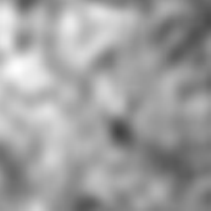}
        \caption*{}
    \end{subfigure}
    \begin{subfigure}{0.19\textwidth}
        \includegraphics[width=\textwidth]{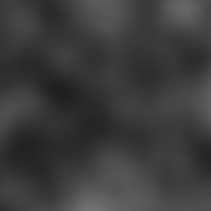}
        \caption*{}
    \end{subfigure}
    \begin{subfigure}{0.19\textwidth}
        \includegraphics[width=\textwidth]{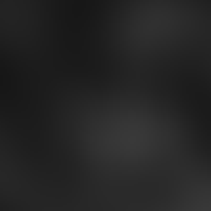}
        \caption*{}
    \end{subfigure}
    \par
    \begin{subfigure}{0.19\textwidth}
        \includegraphics[width=\textwidth]{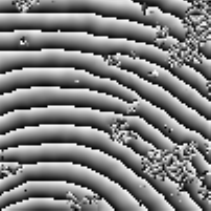}
        \caption*{}
    \end{subfigure}
    \begin{subfigure}{0.19\textwidth}
        \includegraphics[width=\textwidth]{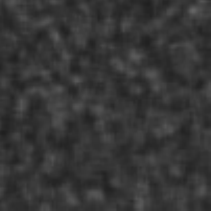}
        \caption*{}
    \end{subfigure}
    \begin{subfigure}{0.19\textwidth}
        \includegraphics[width=\textwidth]{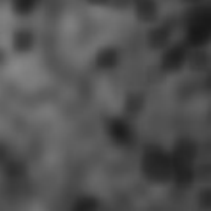}
        \caption*{}
    \end{subfigure}
    \begin{subfigure}{0.19\textwidth}
        \includegraphics[width=\textwidth]{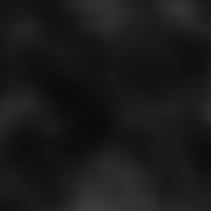}
        \caption*{}
    \end{subfigure}
    \begin{subfigure}{0.19\textwidth}
        \includegraphics[width=\textwidth]{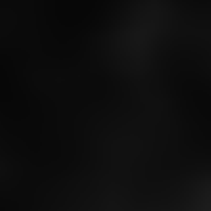}
        \caption*{}
    \end{subfigure}
    \caption{Top row: noisy fingerprint and decomposition into 4 scales. Second row:
    multiscale phase computed via $\mathcal{Q}^{(k)}_{\theta}(\bx)$, local orientation variance quality of each scale.
    Third row: the multiscale phase computed via $A^{(k)}(\bx)$, 
    local amplitude of each scale. Fourth row: the multiscale phase of the product 
    $\mathcal{Q}^{(k)}_{\theta}(\bx) \cdot A^{(k)}(\bx)$, product quality at each scale}
    \label{fig:fingerprint_multiscale_quality_scores}
\end{figure}

\begin{figure}[ht] 
    \centering
    \begin{subfigure}{0.18\textwidth}
        \includegraphics[width=\textwidth]{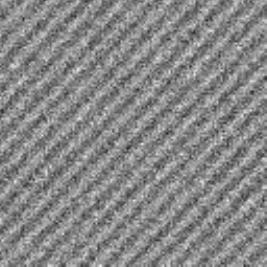}
        \caption{}
    \end{subfigure}%
    \begin{subfigure}{0.18\textwidth}
        \includegraphics[width=\textwidth]{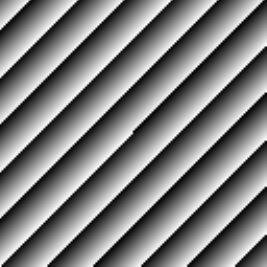}
        \caption{}
    \end{subfigure}
    \begin{subfigure}{0.18\textwidth}
        \includegraphics[width=\textwidth]{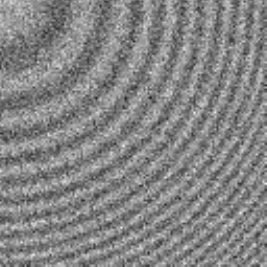}
        \caption{}
    \end{subfigure}%
        \begin{subfigure}{0.18\textwidth}
        \includegraphics[width=\textwidth]{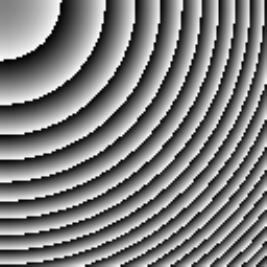}
        \caption{}
    \end{subfigure}%

    \caption{Example of a (a) noisy plane wave and (b) the true underlying phase,
    and (c) a parabolic chirp signal along with (d) its true phase}
    \label{fig:synthetic_examples}
\end{figure}

We propose applying local variance filter $\mathcal{V}$  to each $\theta^{k}$ with an appropriate 
window size, where lower local variance indicates a more coherent signal. 
Define the \textit{local orientation variance quality} and 
corresponding local scale to be:
\begin{equation*}
    \mathcal{Q}^{(k)}_{\theta}(\bx) = \frac{1}{1 + \mathcal{V}_{w_k}(\theta^{k})(\bx)}, 
    \quad k^{\mathcal{Q_{\theta}}}(\bx) = \arg \max_{k} \{\mathcal{Q}^{(k)}_{\theta}(\bx)\}.
\end{equation*} 
We set the window size $w_k$ to be twice the dyadic scale. 
We also consider the \textit{local product quality} given by the product of the local amplitude quality 
and local orientation variance quality, 
$\mathcal{Q}^{(k)}_{\theta}(\bx) \cdot A^{(k)}(\bx)$, as this utilizes both local energy and local structure information.
Figure \ref{fig:fingerprint_multiscale_quality_scores}
compares the results of these three quality functions for the spatial phase estimate of a real fingerprint. 
We further compare phase estimation results of a plane wave signal and a parabolic chirp signal, 
which are shown alongside their respective spatial phases in Figure \ref{fig:synthetic_examples}. 
The details of our experiments are outlined further in Section \ref{sec:4_experiments}.

\section{Experiments}\label{sec:4_experiments}
This section provides results of the proposed multiscale phase estimation algorithm applied 
to several standard phase estimation problems: 1) a baseline experiment on estimating 
the phase of plane waves of different frequences; 2) estimating the phase of a parabolic 
chirp, a standard signal with varying local frequency, which provides a more challenging 
multiscale phase estimation problem; 3) 2D phase demodulation; and
4) a practical example of phase demodulation as it applies to fine-scale fingerprint
registration for the problem of fingerprint matching.

\subsection{Multiscale Phase Estimation Experiments}
\begin{figure}[ht] 
    \centering
    \begin{subfigure}{0.45\textwidth}
        \includegraphics[width=\textwidth,trim={0 0cm 0 0cm},clip]{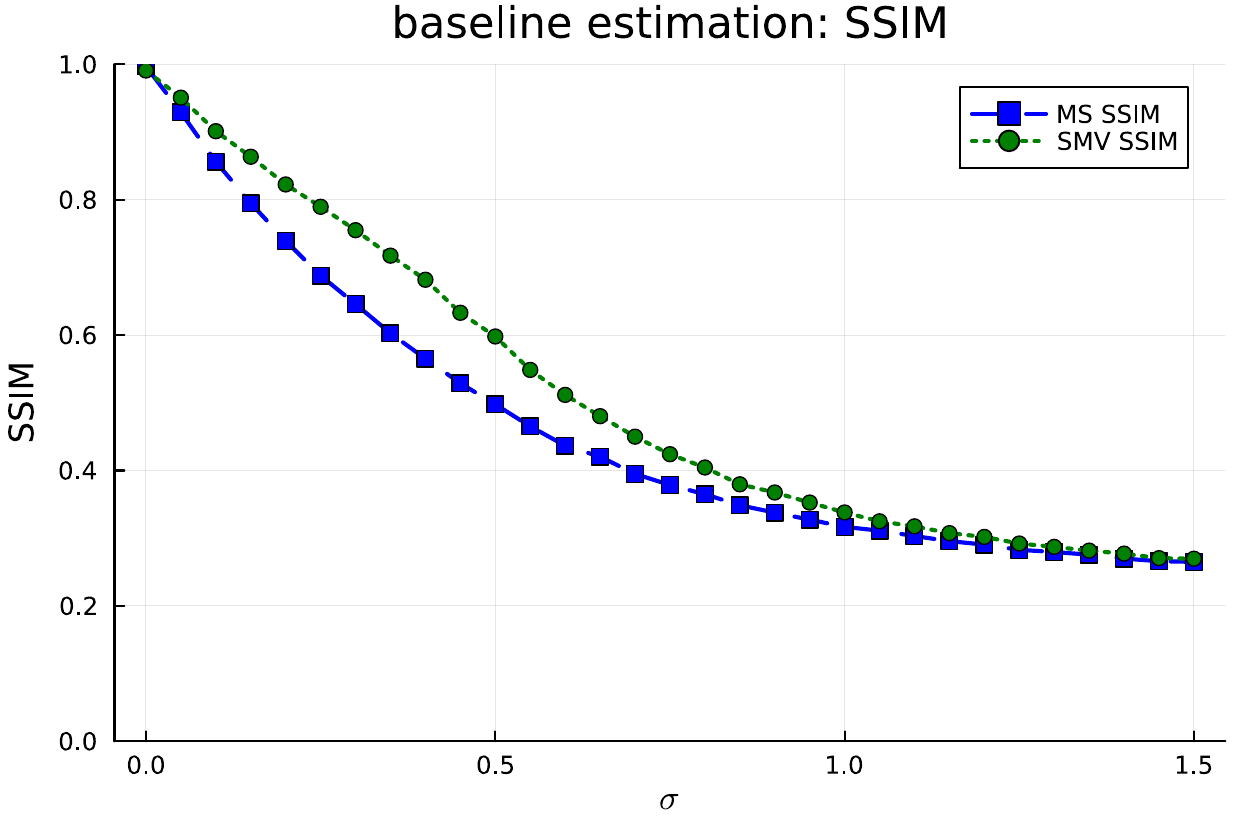}
        \caption{}
    \end{subfigure}%
    \begin{subfigure}{0.45\textwidth}
        \includegraphics[width=\textwidth, trim={0 0cm 0 0cm},clip]{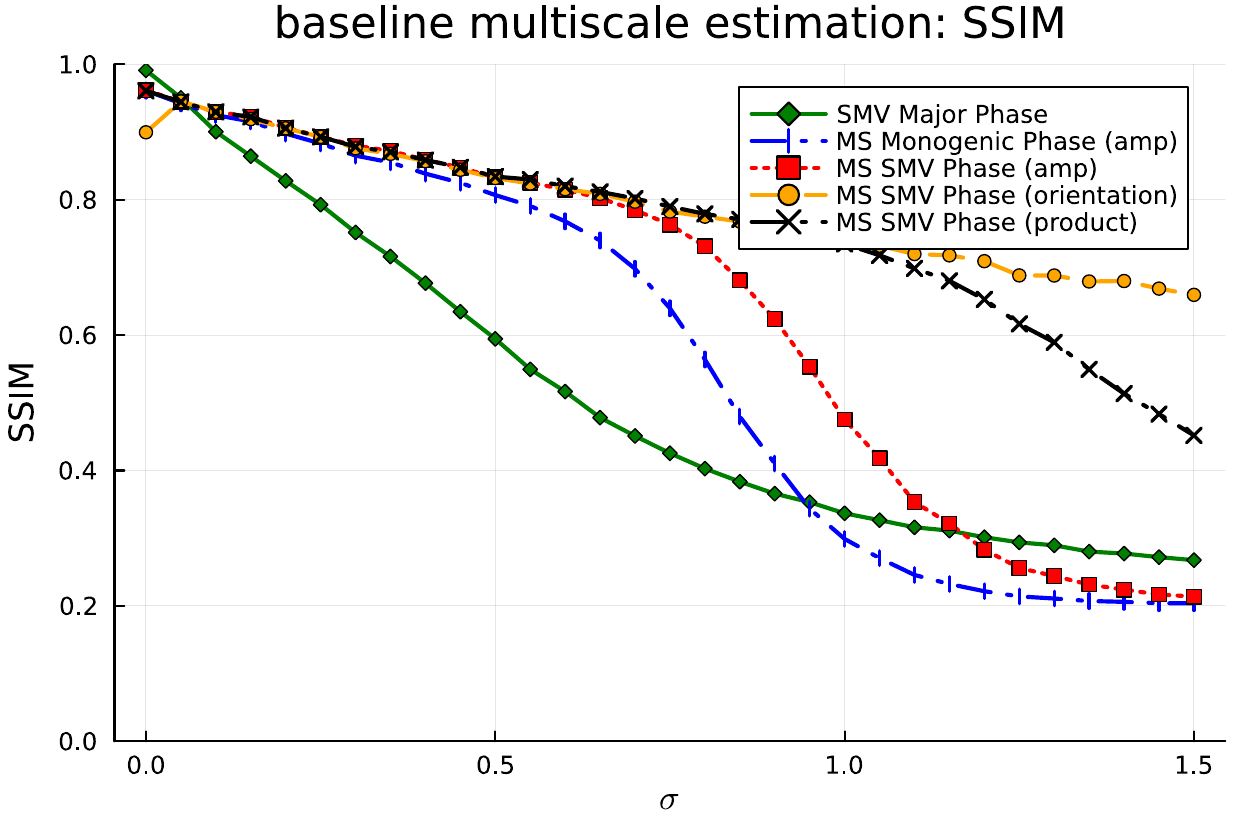}
        \caption{}
    \end{subfigure}
    \caption{phase estimation of noisy plane wave (a) comparison of MS phase vs. SMV phase
    (b) comparison of four multiscale phase estimates}
    \label{fig:baseline_mulstiscale_estimation}
\end{figure}

Given a noisy plane wave of the form 
$f(\bx) = \cos(\omega (\bn \cdot \bx) ) + \eta_\sigma(\bx)$, 
where $\eta_\sigma(\bx)$ is a Gaussian random variable with mean zero and 
standard deviation $\sigma$, the goal is to estimate the true phase function 
$\omega (\bn \cdot \bx) \mod 2\pi$. In our experiment we discretize 
so that $\bx[i,j] = [-\pi + \frac{2\pi i}{N}, -\pi + \frac{2\pi j}{N}]$,
for $i,j = 0, N-1$, for $N = 2^M$, $\omega$ ranges between $2^3$ and $2^{M-2}$,
and $\bn = [\cos(\pi/4), \sin(\pi/4)]^{T}$. 
We use the \emph{structural similary index measure} (SSIM) \cite{1284395} to compare the quality 
of the estimated phase to the ground truth. 
Figure~\ref{fig:baseline_mulstiscale_estimation} shows the quality 
of the estimated phase for $0\le\sigma\le1.5$. Because 
the true amplitude of the noiseless signal is 1 everywhere, 
$\sigma$ can be thought of as the recipricol of SNR, 
hence $\sigma = 1$ is the point at which the Gaussian noise 
begins to dominate the underlying plane wave structure. 
The amplitude-based multiscale phase of the monogenic signal and SMV 
perform well until $\sigma$ surpasses 0.75, beyond which 
the estimate is unusable. In contrast, the estimate from the 
local orientation variance quality and the estimate given by the product of 
the amplitude and local orientation variance quality provide high-quality 
phase estimates well beyond this point. 
Additionally, we test our phase estimation procedure on a \textit{parabolic chirp}. 
The appeal of this signal is that the spatial frequency varies locally, 
and so it is a more challenging phase estimation task.
Estimating the phase of such a signal is actually a known strength of the monogenic signal \cite{kaseb_phase_2019}.
We demonstrate again that in the presence of signal corruption the monogenic signal fails quickly, but our multiscale 
approach finds the coherent local structure reliably.
Furthermore, because the monogenic signal (SMV) handles locally varying frequency well, we posit using an 
``overcomplete'' set of features can improve phase estimation. 
We use a set of features which includes the low-pass component at each dyadic scale. 

\begin{figure}[ht] 
    \centering
    \begin{subfigure}{0.45\textwidth}
        \includegraphics[width=\textwidth]{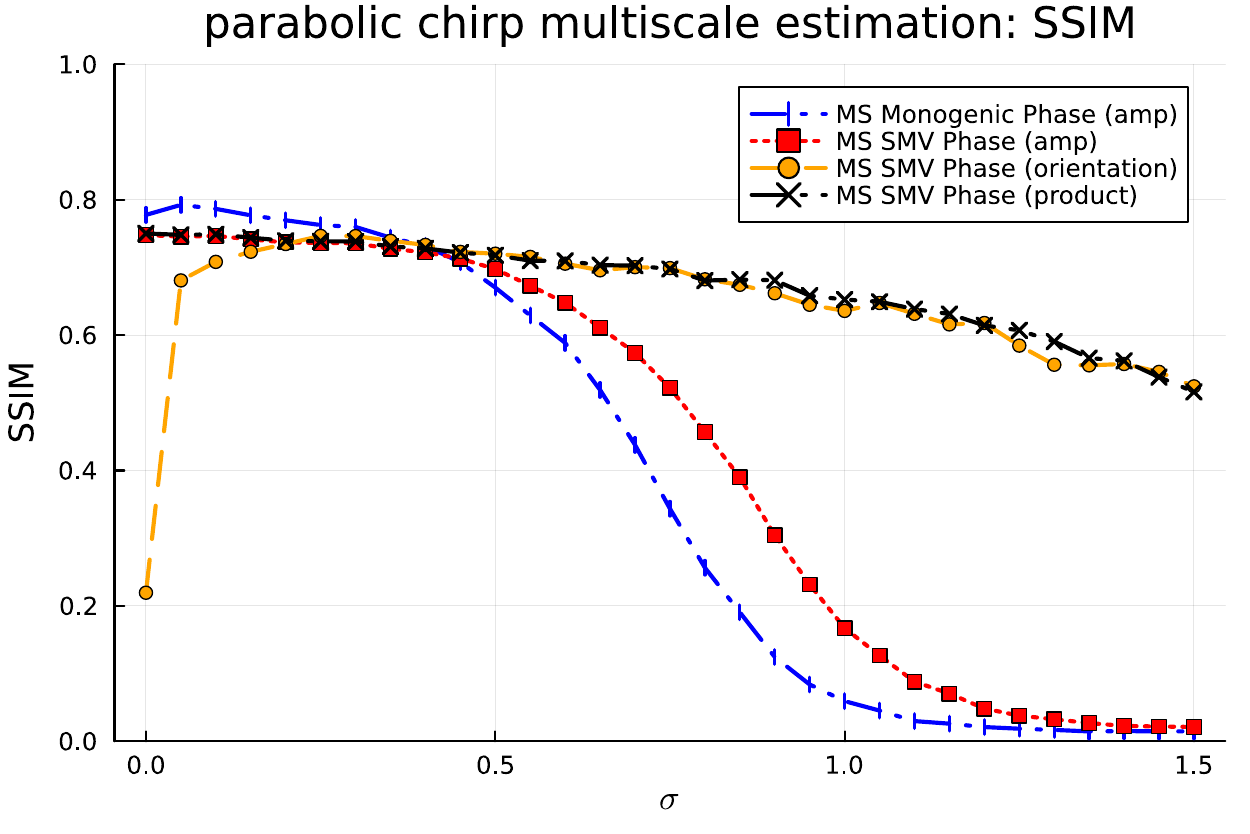}
        \caption{}
    \end{subfigure}%
    \begin{subfigure}{0.45\textwidth}
        \includegraphics[width=\textwidth]{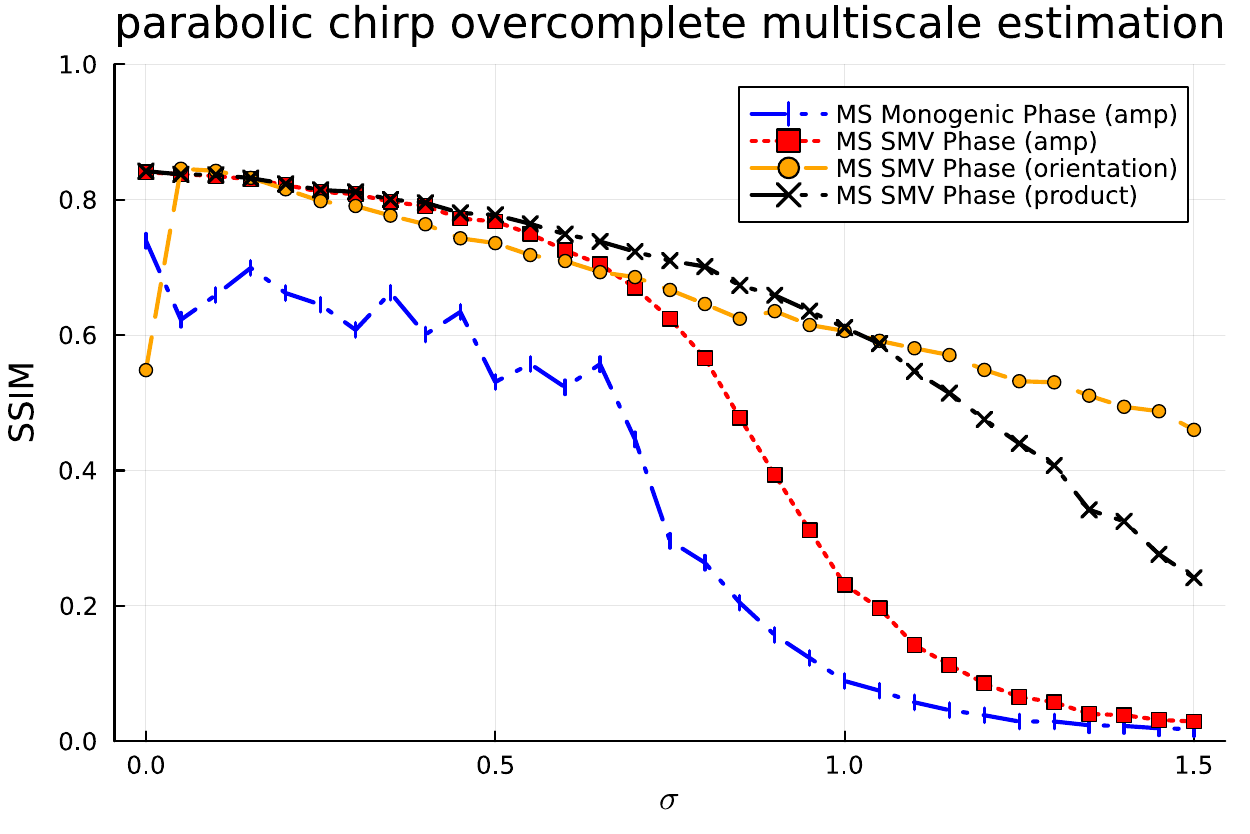}
        \caption{}
    \end{subfigure}
    \caption{Phase estimation of noisy parabolic chirp (a) comparison of four multiscale phase 
    estimates
    (b) comparison of same multiscale estimation procecures with an overcomplete set 
    of scales}
    \label{fig:parabolic_chirp_multiscale_estimation}
\end{figure}

Figure~\ref{fig:parabolic_chirp_multiscale_estimation} shows the results of the 
multiscale estimate and the overcomplete multiscale estimate for varying SNR.

\subsection{Phase Demodulation with Multiscale Major Phase}
\begin{figure} 
    \centering
    \begin{subfigure}{0.5\textwidth}
    \centering
        \begin{subfigure}{0.30\textwidth}
            \renewcommand\thesubfigure{\alph{subfigure}1}
            \includegraphics[width=\textwidth]{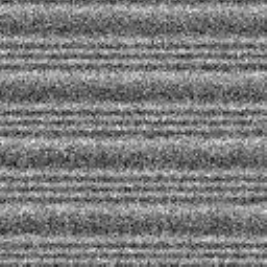}
            \centering
            \caption{}
        \end{subfigure}
        \begin{subfigure}{0.30\textwidth}
            \addtocounter{subfigure}{-1}
            \renewcommand\thesubfigure{\alph{subfigure}2}
            \centering
            \includegraphics[width=\textwidth]{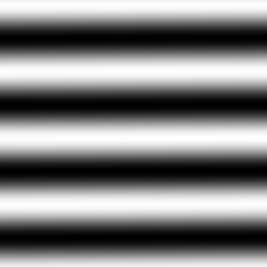}
            \caption{}
        \end{subfigure}
        \addtocounter{subfigure}{-1}
        \par
        \begin{subfigure}{0.30\textwidth}
            \renewcommand\thesubfigure{\alph{subfigure}3}
            \includegraphics[width=\textwidth]{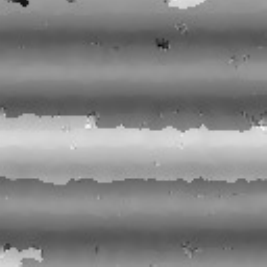}
            \centering
            \caption{}
        \end{subfigure}
        \begin{subfigure}{0.30\textwidth}
            \addtocounter{subfigure}{-1}
            \renewcommand\thesubfigure{\alph{subfigure}4}
            \centering
            \includegraphics[width=\textwidth]{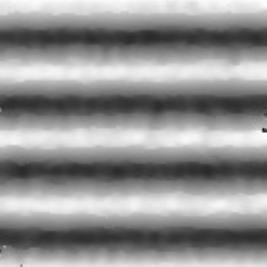}
            \caption{}
        \end{subfigure}
    \end{subfigure}%
    \begin{subfigure}{0.5\textwidth}
        \centering
        \includegraphics[width=\textwidth]{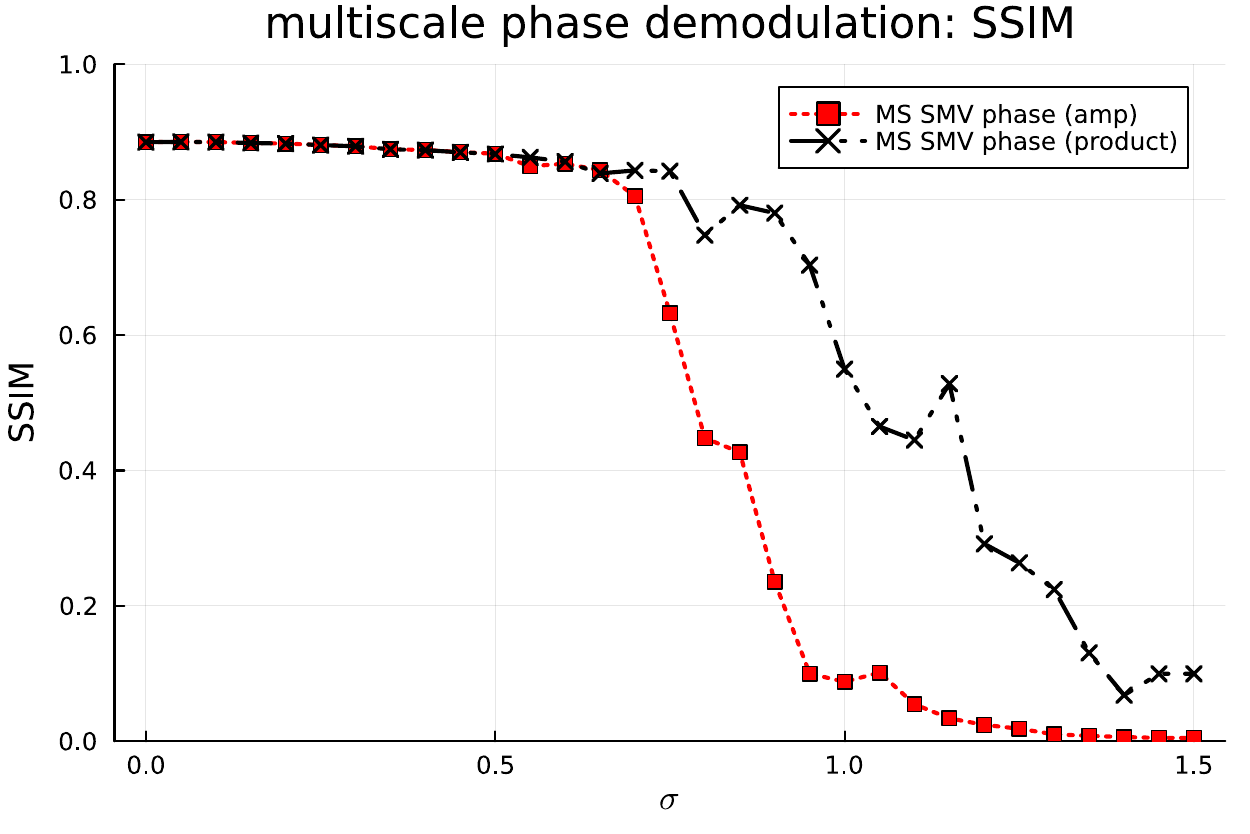}
        \caption{}
    \end{subfigure}
    \caption{Phase demodulation results: (a1) noisy phase modulated message, $\sigma = 0.75$, 
    (a2) ground truth message, (a3) recovered via MS SMV amplitude quality, (a4) recovered via MS SMV product quality
    (b) comparison of MS SMV phase estimate using the amplitude quality (square markers) versus product quality (x markers)}
    \label{fig:phase_demodulation_results}
\end{figure}

The multiscale major phase estimate given by the the multiscale SMV representation of a signal $f$ motivates 
further experiments in phase demodulation. In two dimensions, the phase
demodulation problem can be stated as follows: 
Given the carrier wave $c(\bx) = A\cos(\omega_c\bn \cdot \bx + \phi_c)$, and a message $m(\bx)$
where $\hat{m}(\bu) = 0$ for $|\bu| \ge \omega_c$, the phase modulated (PM) signal is given by:
\begin{equation}
    c_{PM}(\bx) = \cos(\omega_c\bn \cdot \bx + \phi_c + m(\bx)).
\end{equation}
Typically the message is also sinusoidal in structure.
For a baseline phase demodulation task, we attempt to recover a sinusoidal message after phase modulation
and noise corruption. In this case, motivated by the phase estimation of the parabolic chirp, we 
use the overcomplete feature set when computing the multiscale phase to more accurately capture the 
overlapping frequency bands resulting from the phase modulation. 
After estimating the modulated phase, $\tilde\Phi$, it is unwrapped to produce 
$\tilde{\Phi}^{u}$. If successful, the message should be well estimated by either 
$\pm\tilde{\Phi}^{u} - \omega_c \bn \cdot \bx$. 
The $\pm$ here is due to sign ambiguity in the phase demodulation problem; 
for sythetic experiments the ground truth message allows us to choose the proper sign.
Figure~\ref{fig:phase_demodulation_results} shows the noisy phase modulated message,
the ground truth message, and recovered messages via an amplitude based 
mulstiscale phase estimate versus the proposed product method. 
Again our tests conclude the product quality map allows for accurate phase 
estimation even when SNR is large.

\subsection{Application to Deformable Fingerprint Registration}
In 2018 Cui et al \cite{cui_Deformable_8368301} proposed a method for 
fine-scale fingerprint registration via phase demodulation. 
The method is as follows, given a fixed and moving image, $f_{f}$ and $f_{m}$,
which have already been coarsely registered, we consider $f_{f}$ to be 
the carrier wave and $f_{m}$ to be a phase-modulated signal, where the 
message represents the unknown displacement vector field.
Let $T(\bx)$ represent this displacement vector field between $f_{f}(\bx)$
and $f_{m}(\bx)$ such that  $f_{f}(\bx + T(\bx)) = f_{m}(\bx)$. 
Let $\bx' = \bx + T(\bx)$.
Compute $\phi_{f}$ and $\phi_{m}$, the (wrapped) spatial phase of the fixed and moving image 
respectively, and let $\Delta \phi = \phi_{f} - \phi_{m}$. Then $\Delta \phi^u$, the
unwrapped phase differences, combined with local frequency information
can be used to compute a spatial displacement
at each coordinate, which enable the fine-scale deformable registration step. 
\begin{figure}[ht] 
    \centering
    \begin{subfigure}{0.495\textwidth}
    \centering
        \begin{subfigure}{0.25\textwidth}
            \renewcommand\thesubfigure{\alph{subfigure}1}
            \includegraphics[width=\textwidth]{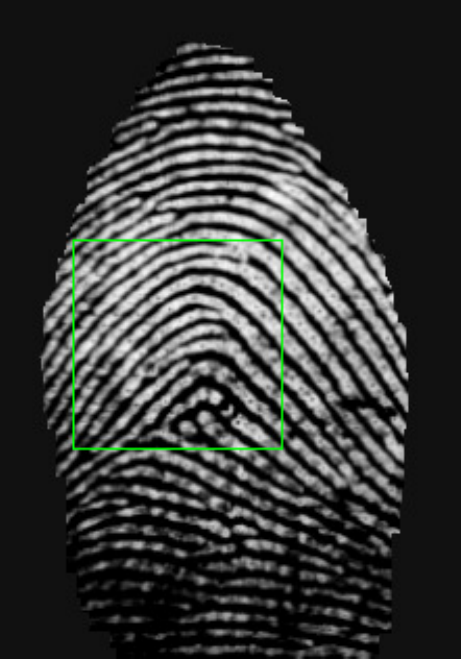}
            \centering
            \caption{}
        \end{subfigure}
        \begin{subfigure}{0.25\textwidth}
            \addtocounter{subfigure}{-1}
            \renewcommand\thesubfigure{\alph{subfigure}2}
            \centering
            \includegraphics[width=\textwidth]{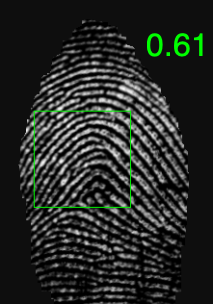}
            \caption{}
        \end{subfigure}
        \begin{subfigure}{0.25\textwidth}
            \addtocounter{subfigure}{-1}
            \renewcommand\thesubfigure{\alph{subfigure}3}
            \centering
            \includegraphics[width=\textwidth]{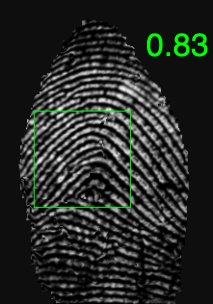}
            \caption{}
        \end{subfigure}
        \addtocounter{subfigure}{-1}
        \begin{subfigure}{0.25\textwidth}
            \renewcommand\thesubfigure{\alph{subfigure}4}
            \includegraphics[width=\textwidth]{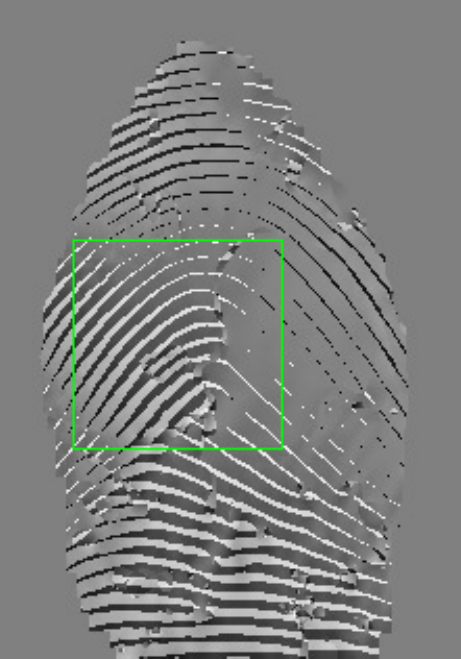}
            \centering
            \caption{}
        \end{subfigure}
        \begin{subfigure}{0.25\textwidth}
            \addtocounter{subfigure}{-1}
            \renewcommand\thesubfigure{\alph{subfigure}5}
            \centering
            \includegraphics[width=\textwidth]{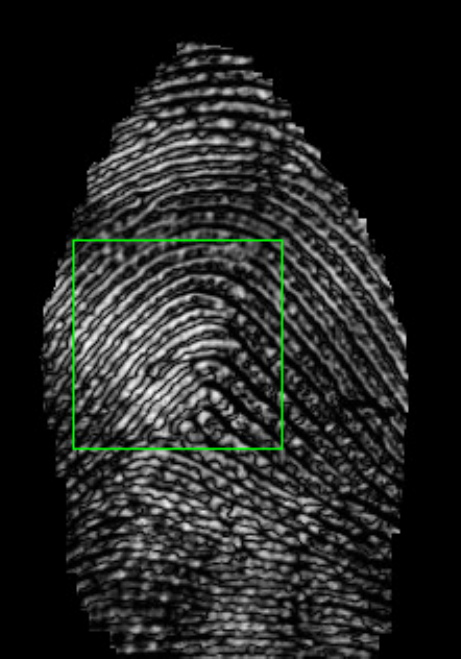}
            \caption{}
        \end{subfigure}
        \begin{subfigure}{0.25\textwidth}
            \addtocounter{subfigure}{-1}
            \renewcommand\thesubfigure{\alph{subfigure}6}
            \centering
            \includegraphics[width=\textwidth]{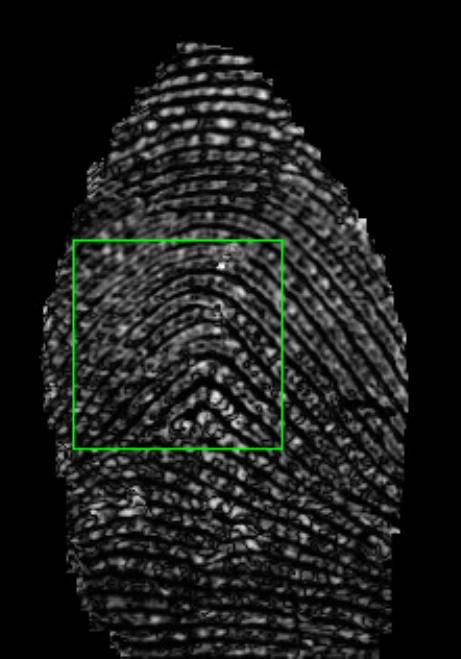}
            \caption{}
        \end{subfigure}
    \end{subfigure}
    \begin{subfigure}{0.495\textwidth}
        \setlength{\belowcaptionskip}{0.5\baselineskip}
        \centering
        \begin{subfigure}{0.3\textwidth}
            \renewcommand\thesubfigure{\alph{subfigure}1}
            \includegraphics[width=\textwidth]{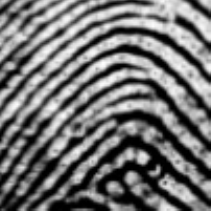}
            \centering
            \caption{}
        \end{subfigure}
        \begin{subfigure}{0.3\textwidth}
            \addtocounter{subfigure}{-1}
            \renewcommand\thesubfigure{\alph{subfigure}2}
            \centering
            \includegraphics[width=\textwidth]{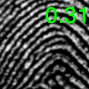}
            \caption{}
        \end{subfigure}
        \begin{subfigure}{0.3\textwidth}
            \addtocounter{subfigure}{-1}
            \renewcommand\thesubfigure{\alph{subfigure}3}
            \centering
            \includegraphics[width=\textwidth]{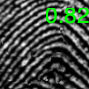}
            \caption{}
        \end{subfigure}
        \addtocounter{subfigure}{-1}
        \begin{subfigure}{0.3\textwidth}
            \renewcommand\thesubfigure{\alph{subfigure}4}
            \includegraphics[width=\textwidth]{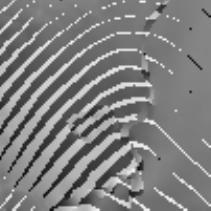}
            \centering
            \caption{}
        \end{subfigure}
        \begin{subfigure}{0.3\textwidth}
            \addtocounter{subfigure}{-1}
            \renewcommand\thesubfigure{\alph{subfigure}5}
            \centering
            \includegraphics[width=\textwidth]{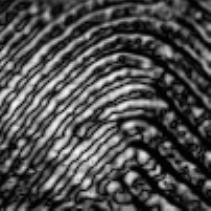}
            \caption{}
        \end{subfigure}
        \begin{subfigure}{0.3\textwidth}
            \addtocounter{subfigure}{-1}
            \renewcommand\thesubfigure{\alph{subfigure}6}
            \centering
            \includegraphics[width=\textwidth]{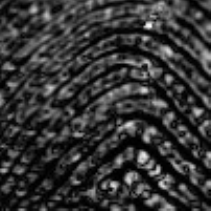}
            \caption{}
        \end{subfigure}
    \end{subfigure}%
    \caption{(a1) fixed image, (a2) moving image, (a3) registered image,  
    (a4) difference of multiscale phases, (a5) $|f_m(\bx) - f_f(\bx)|$ (a6) $|f_r(\bx) - f_f(\bx)|$. 
    (b1) fixed image, (b2) moving image (b3) registered image,
    (b4) difference of multiscale phases, (b5) $|f_m(\bx) - f_f(\bx)|$ (b6) $|f_r(\bx) - f_f(\bx)|$.
    The overlaid numbers indicate the correlation coefficient with respect to the fixed image.}
    \label{fig:fingerprint_deformation_example}
\end{figure}
More explicitly, we let $f_{f}(\bx) = \cos(\phi_{f}^u(\bx))$ 
and $f_{m}(\bx) = \cos(\phi_{m}^u(\bx))$ and assume that near $\bx = \bx_0$ we have the estimate 
$\phi(\bx) = 2\pi \omega \bn \cdot \bx.$ 
Then $f_m(\bx) = f_f(\bx') = \cos(2\pi \omega \bn \cdot \bx + 2\pi\omega\bn\cdot T(\bx))$ and 
$\Delta\phi^u(\bx) = 2\pi\omega\bn\cdot T(\bx)$, so we recover $\bn \cdot T(\bx) = \frac{\Delta\phi^u(\bx)}{2\pi\omega}$,
which gives the magnitude of the projection of $T(\bx)$ along the local orientation $\bn = [\cos(\theta) \sin(\theta)]^T$. 
If we restrict our displacement along this direction, then, we have
\begin{equation}
        d_x(\bx) = \frac{\Delta\phi^u(\bx)}{2\pi\omega}\cos(\theta), \quad
        d_y(\bx) = \frac{\Delta\phi^u(\bx)}{2\pi\omega}\sin(\theta).
\end{equation}
When we have estimates of the local phase, orientation, and frequency, this becomes
\begin{equation}
        d_x(\bx) = \frac{\Delta\phi^u(\bx)}{2\pi\omega(\bx)}\cos(\theta(\bx)),\quad
        d_y(\bx) = \frac{\Delta\phi^u(\bx)}{2\pi\omega(\bx)}\sin(\theta(\bx)).
\end{equation}
Global phase unwrapping algorithms typically have several major discontinuities which provide unreliable frequency information.
In our experiments we compute a local frequency estimate $\omega(\bx)$ by differentiation 
of a windowed phase unwrapping of $\Delta\phi(\bx)$. 
The phase and orientation values used are those provided by our multiscale method. 
Figure~\ref{fig:fingerprint_deformation_example} gives an example of 
the fine-scale registration produced by this method 
(all images from FVC2004 DB1-B \cite{10.1007/978-3-540-25948-0_1}).
\begin{table}[ht]
    \vspace{-1em}
    \centering
    \caption{Correlation Values before and after fine-scale registration for various
    noise levels.}\label{table:fingerprint_assessment}
    \begin{tabular}{|c|c|c|c|}
    \hline
    Type &  Affine & Fine-scale& Difference\\
    \hline
    $\sigma=0.0$ & 0.76 & 0.83 & 0.070\\
    $\sigma=0.1$ & 0.72 & 0.80 & 0.076\\
    $\sigma=0.2$ & 0.62 & 0.71 & 0.086\\
    $\sigma=0.3$ & 0.51 & 0.60 & 0.091\\
    $\sigma=0.4$ & 0.41 & 0.50 & 0.085\\
    $\sigma=0.5$ & 0.33 & 0.40 & 0.070\\
    \hline
    \end{tabular}
    \vspace{-1em}
\end{table}

Table \ref{table:fingerprint_assessment} expresses the quality of the fingerprint match after 
successful rigid registration of fingerprints, and then after the additional fine-scale deformable registration is applied.
Because we are interested only in the performance of the fine-scale registration algorithm,
we add noise only after successful rigid registration. 
We find that even at large noise levels we are able to estimate accurate fine-scale registration in 
well structured areas of the fingerprints.

\section{Conclusion}\label{sec:5_conclusion}
We have argued that the Structure Multivector is a robust method for estimating
the local energy and structure of fringe and interference patterns, and further define a local quality metric 
which rewards areas of coherent local structure. 
The result is a robust spatial phase estimation algorithm which allows for accurate spatial phase estimation 
even as noise begins to dominate the signal. We have demonstrated this with several synthetic examples and in 
a practical setting of fine-scale fingerprint registration.

\begin{credits}
\subsubsection{\ackname}
This research was partially supported by the US National Science Foundation grantsDMS-1912747 
and CCF-1934568 as well as the US Office of Naval Research grant N00014-20-1-2381.
\subsubsection{\discintname}
The authors have no competing interests to declare that are
relevant to the content of this article.
\end{credits}
%
%
\bibliographystyle{splncs04}
\bibliography{refs}
\pagebreak
\appendix

\section{Construction of the SMV}\label{sec:appendixA_SMV}
Two-dimensional signals can, of course, vary in two or more orientations in any given 
local patch, and further work has been done using hypercomplex signal processing that can deal with these cases. One extension, the \textit{structure multivector} (SMV), 
was introduced along with the monogenic signal in the PhD dissertation of Felsberg
\cite{felsberg_low-level_nodate}. It is designed to deal with signals of the form
\begin{equation}\label{eq:i2DSignal_appendix}
    f(\bx) = f_1(\bn(\bx)\cdot\bx) + f_2(\bn(\bx)^{\perp} \cdot \bx),
\end{equation}
These are intrinsically 2D (i2D) signals with two orientations in each local patch that are 
orthogonal to one another.

The features of the SMV will essentially be that of two monogenic signals, and to accommodate these 
additional features the SMV lives in a larger dimensional Clifford algebra, $\Cl{3}$ which 
subsumes the quaternions. 
For more information on Clifford algebras see \cite{delanghe_clifford_2001}, \cite{felsberg_low-level_nodate}. 
In general we denote the product 
$e_{i_{1}}e_{i_{2}}\cdots e_{i_{n}} := e_{i_{1}i_{2} \cdots i_{n}}$.

We consider an image of the form $\bff:\Rf^2 \to e_3\Rf$,
$$
    \bff(xe_1 + ye_2) = f(x,y)e_3.
$$
Note $f = \bff e_3 = e_3 \bff$.

The corresponding structure multivector (SMV) is given by 

\begin{equation*}\label{eq:SMV_appendix}
    \begin{split}
        M_S(\bx) &=  \left[\bff(\bx) + (h_2^1 * \bff)(\bx)\right] + 
        e_3 \left[(h_2^2*\bff)(\bx) + (h_2^3 * \bff)(\bx)\right]\\
        &= M_0 + M_1 e_1 + M_2 e_2 + M_3 e_3 + 
        M_{23}e_{23} + M_{31} e_{31} + M_{12} e_{12}
    \end{split}
\end{equation*}

The explicit definitions of these functions are given below, 
where $\bx = x e_1 + y e_2$:
\begin{align*}
        M_{1} &= \frac{x}{2\pi|\bx|^3} * f(\bx), &
        M_{2} &= \frac{y}{2\pi|\bx|^3} * f(\bx), &
        M_{3} &= f(\bx),\\
        M_{23} &= \frac{3(3x^2y - y^3)}{2\pi|\bx|^5}*f(\bx), &
        M_{31} &= \frac{3(3xy^2 - x^3)}{2\pi|\bx|^5}*f(\bx), &&\\
        M_{0} &= \frac{-2(x^2 - y^2)}{2\pi|\bx|^4} * f(\bx), &
        M_{12} &= \frac{-4xy}{2\pi|\bx|^4} * f(\bx). &&
\end{align*}

The Fourier transforms of the operators $h^k_{2}$ are given below:
\begin{equation*}
    \begin{split}
    H_2^1(\bu) &= \frac{\bu}{|\bu|}I_2^{-1} = \frac{-ue_2 + ve_1}{|\bu|}\\
    H_2^2(\bu) &= \frac{e_1\bu e_1\bu}{\bu^2} = 
    \frac{(u^2 - v^2) + 2uv e_{12}}{\bu^2}\\
    H_2^3(\bu) &= \frac{\bu e_1 \bu e_1\bu}{|\bu|^3}I_2^{-1} = 
    \frac{(3uv^2 - u^3)e_{23} - (3u^2v - v^3)e_{31}}{|\bu|^3},
    \end{split}
\end{equation*}
where $I_2 = e_{12}$ and $I_2^{-1} = e_{21}$ as $e_{12} e_{21} = 1$ by 
definition. $I_2$ acts an imaginary unit here, as $I_2^2 = -1$.
Note $h_{2}^1$ is simply the Riesz transform, and $h_2^3$ is the composition of $h_2^2$ and the Riesz transform, 
so the crux of this extension is in understanding $H_2^2$. First, it responds only to even signals. 
Second, any two perpendicular vectors $\bn$ and $\bn^{\perp}$ are antiparallel after action by $H_2^2$, 
which means that an even signal according to the \eqref{eq:i2DSignal_appendix} will yield a response to $H^2_2$ 
whose argument is precisely twice that of the main orientation of $\bn$. 

Specifically, we can calculate the orientation $\bn$ given a signal of 
the form $f(\bx) = A\cos(\bn\cdot\bx) + B\cos(\bn^{\perp}\cdot\bx)$ directly from this response. 
To handle odd structures, we finally take the Riesz transform of $h_2^2$ to yield $h_3^2$. 
The product of the Riesz response and the response of the third order harmonic estimates this same orientation,
but is better suited for odd structures, hence the average of these two
arguments provides a robust orientation estimate of the structure multivector, 
as given in Felsberg's dissertation \cite{felsberg_low-level_nodate}:

\begin{equation}\label{eq:SMVOrientationEst_appendix}
    \theta_e = \frac14 \arg \left[(M_0 + M_{12} I_2)^2 + (M_1 + M_2 I_2)(M_{31} - M_{23}I_2)\right].
\end{equation}

In \cite{felsberg_low-level_nodate} he shows that the extended signal model provides a more robust orientation estimator 
than that of the monogenic signal. This is further confirmed in \cite{sinusoidal_image_model}. 
In addition to these facts, we show that: 1) the feature set of the SMV is robust even to i2D signals which violate 
the orthogonality constraint; and 2) if one of the local i1D signal dominates the local energy, 
then we can estimate the corresponding orientation well even in the case of large deviation from this 
constraint. See Appendix~\ref{sec:appendixB_orientation} for details.

With this orientation estimate it is then possible to construct 
a pair of angular filters that decompose a signal $f$ into two i1D signals, 
which will then yield two local amplitudes, two local orientations, 
and two local phases that can be used for further processing.

Explicitly, these filters are given by:
\begin{equation*}
    \begin{split}
        \mathcal{W}_1f &= \frac{M_3 + \cos(2\theta_e)M_0 + \sin(2\theta_e)M_{12}}{2}\\
        \mathcal{W}_2f &= \frac{M_3 - \cos(2\theta_e)M_0 - \sin(2\theta_e)M_{12}}{2}
    \end{split}
\end{equation*}
and their Riesz transforms, projected onto $-e_2$:
\begin{equation*}
    \begin{split}
        \mathcal{W}_3f &= \frac{3(\cos(\theta_e)M_1 + \sin(\theta_2)M_2) + \cos(3\theta_e)M_{31} - 
        \sin(3\theta_e)M_{23}}{4}I_2\\
        \mathcal{W}_4f &= \frac{3(-\sin(\theta_e)M_1 + \cos(\theta_2)M_2) + \sin(3\theta_e)M_{31} + 
        \cos(3\theta_e)M_{23}}{4}I_2.\\
    \end{split}
\end{equation*}

Then our two complex i1D signals are given by 
\begin{equation*}
        F_1(\bx) = \mathcal{W}_1f + \mathcal{W}_3f,\quad
        F_2(\bx) = \mathcal{W}_2f + \mathcal{W}_4f
\end{equation*}

Recall our orientation $\theta_e$ is a function of the spatial variable $\bx$ and is implicit in the above signals; 
see Eq.~\eqref{eq:SMVOrientationEst} (or Eq.~\eqref{eq:SMVOrientationEst_appendix}).

The full feature set of the SMV then is given by this local orientation estimate and:
\begin{equation*}
        A_i(\bx) =  |F_i(\bx)|,\quad \phi_i(\bx) = \arg|F_i(\bx)|,
\end{equation*}
for $i = 1, 2$. At each location $\bx$, we choose the main signal by selecting the pair with the largest local amplitude. 
This selection is given by the dominance index $d(\bx) = \argmax_{1,2} \{A_1(\bx), A_2(\bx)\}$, 
so that we have a major and minor IAP representation given by:
\begin{align*}
    A(\bx) &=  A_{d(\bx)}, \qquad \Phi(\bx) = \phi_{d(\bx)},\\
    a(\bx) &=  A_{3 - d(\bx)}, \quad \phi(\bx) = \phi_{3 - d(\bx)}.
\end{align*}
Here the capital $A$ and $\Phi$ denote the dominant, or major, local i1D signal. 

\section{Orientation Estimation with the SMV}\label{sec:appendixB_orientation}

Our signal model \eqref{eq:i2DSignal} (also \eqref{eq:i2DSignal_appendix}) assumes the sum of two orthogonal 
sinusoidal modes, and we would like to know how this orientation estimate 
behaves if they are not perfectly orthogonal. We will restrict ourselves 
to the case where they have equal frequency so that they are 
inseparable even in the multiscale phase model. 

Suppose $\bn = \cos(\theta)e_1 + \sin(\theta)e_2$,
$\bn^{\perp}_{\epsilon} = \sin(\theta+\epsilon)e_1 - \cos(\theta+\epsilon)e_2$,
and 
$$
    f(\bx) = A\cos(\bn\cdot\bx) + B\cos(\bn^{\perp}_{\epsilon}).
$$

We have the following:
\begin{equation*}
    \begin{split}
        M_1 + M_2I_2 &= A (e_1\bn)\sin(\bn \cdot \bx) + B e_1\bn^{\perp}_{\epsilon}\sin(\bn^{\perp}_{\epsilon} \cdot \bx)\\
        M_0 + M_{12}I_2 &= A (e_1\bn)^2\cos(\bn \cdot \bx) + 
        B (e_1\bn^{\perp}_{\epsilon})^2\cos(\bn^{\perp}_{\epsilon} \cdot \bx)\\
        M_{31} - M_{23}I_2 &= A (e_1\bn)^3\sin(\bn \cdot \bx) + 
        B (e_1\bn^{\perp}_{\epsilon})^3\sin(\bn^{\perp}_{\epsilon} \cdot \bx),
    \end{split}
\end{equation*}
where 
\begin{equation*}
    \begin{split}
        e_1 \bn^{\perp}_{\epsilon} &= -I_2\exp(\theta I_2)\exp(\epsilon I_2)\\
        &= -I_2\exp(\epsilon I_2) (e_1\bn)\\
        (e_1 \bn^{\perp}_{\epsilon})^2 &= -\exp(2\theta I_2)\exp(2\epsilon I_2)\\
        &= -\exp(2\epsilon I_2)(e_1\bn)^2\\
        (e_1 \bn^{\perp}_{\epsilon})^3 &= I_2\exp(3\theta I_2)\exp(3\epsilon I_2)\\
        &= I_2\exp(3\epsilon I_2)(e_1\bn)^3.
    \end{split}
\end{equation*}
Analyzing the even response first, we have
\begin{equation*}
    \begin{split}
    (M_0 + M_{12}I_2)^2 &= (e_1\bn)^4\\
    &[A^2\cos^2(\bn\cdot\bx) + 
    B^2\exp(4\epsilon I_2) \cos^2(\bn^{\perp}_{\epsilon}\cdot\bx) \\
    &+2AB\exp(2\epsilon I_2)\cos(\bn\cdot\bx)\cos(\bn^{\perp}_{\epsilon}\cdot\bx)].
    \end{split}
\end{equation*}
For the product of odd responses we have:
\begin{equation*}
    \begin{split}
        (M_1 + M_{2}I_2)(M_{31} - M_{23}I_2) &= 
        (e_1\bn)^4[A^2\sin^2(\bn\cdot\bx) \\
        &+ B^2\exp(4\epsilon I_2) \sin^2(\bn^{\perp}_{\epsilon}\cdot\bx) + AB(\exp(3\epsilon I_2) \\
        &+ \exp(\epsilon I_2))
        \sin(\bn\cdot\bx)\sin(\bn^{\perp}_{\epsilon}\cdot\bx)].
    \end{split}
\end{equation*}
In general for any real number $a$ we have: $\exp(ai) + \exp(3ai) = 2\exp(2ai)\cos(a)$, 
our odd response becomes
\begin{equation*}
    \begin{split}
        (M_1 + M_{2}I_2)(M_{31} - M_{23}I_2) &= 
        (e_1\bn)^4 [A^2 + B^2\exp(4\epsilon I_2) \\
        &+ 2AB\exp(2\epsilon I_2)[
        \cos(\bn\cdot\bx)\cos(\bn^{\perp}_{\epsilon}\cdot\bx)
        +\\
        &\cos(\epsilon) \sin(\bn\cdot\bx)\sin(\bn^{\perp}_{\epsilon}\cdot\bx)]].
    \end{split}
\end{equation*}
Therefore, letting $M = 
\cos(\bn\cdot\bx)\cos(\bn^{\perp}_{\epsilon}\cdot\bx)
+\cos(\epsilon) \sin(\bn\cdot\bx)\sin(\bn^{\perp}_{\epsilon}\cdot\bx)$ for brevity,
our estimate is give by:
\begin{equation*}
    \begin{split}
        \Theta(\epsilon) &= \theta + \frac14 \arg [A^2 + B^2\exp(4\epsilon I_2) + 
        2AB\exp(2\epsilon I_2)\cdot M]\\
        &= \theta + \frac14 \tan^{-1}\left[
        \frac{B^2\sin(4\epsilon) + 2AB\sin(2\epsilon) \cdot M}
        {A^2 + B^2\cos(4\epsilon) + 2AB\cos(2\epsilon)\cdot M
        }
        \right].
    \end{split}
\end{equation*}

 We see that as $\frac{B}{A}\to 0$ it follows that $\Theta(\epsilon) \to \theta$.

Additionally, in the case $A = B$, the argument in the artan function becomes
\begin{equation*}
    \begin{split}
            \frac{\sin(4\epsilon) + 2\sin(2\epsilon)\cdot M}
            {1 + \cos(4\epsilon) + 2\cos(2\epsilon)\cdot M}
            &=\frac{2\sin(2\epsilon)[\cos(2\epsilon)+ M]}
            {2\cos(2\epsilon)[\cos(2\epsilon) + M]
            }\\
            &= \tan(2\epsilon).
    \end{split}
\end{equation*}
Thus 
$$
    \Theta(\epsilon) = \theta + \frac{\epsilon}{2}.
$$
for $|\epsilon| < \pi/4$.

\end{document}